\documentclass[pss]{wiley2sp} 
\usepackage{amsmath}
\usepackage{siunitx}

\DeclareSIUnit\micron{\micro\metre}
\DeclareSIUnit\nb{\nano\barn}
\DeclareSIUnit\pb{\pico\barn}
\DeclareSIUnit\fb{\femto\barn}
\DeclareSIUnit\particle{p}

\begin{document}

\title{Description of radiation damage in diamond sensors using an effective defect model}

\titlerunning{Diamond polarization}

\author{%
  Florian Kassel\textsuperscript{\Ast,\textsf{\bfseries 1,2}},
  Moritz Guthoff\textsuperscript{\textsf{\bfseries 2}},
  Anne Dabrowski\textsuperscript{\textsf{\bfseries 2}},
  Wim de Boer\textsuperscript{\textsf{\bfseries 1}}}

\authorrunning{Florian Kassel et al.}

\mail{e-mail
  \textsf{florian.kassel@cern.ch}, Phone: +41 75 411 7486}

\institute{%
  \textsuperscript{1}\,Institute for Experimental Nuclear Physics (IEKP), KIT, Karlsruhe, Germany\\
  \textsuperscript{2}\,CERN, Meyrin, Switzerland}

\received{XXXX, revised XXXX, accepted XXXX} 
\published{XXXX} 


\abstract{%
%
%
%
\abstcol{%
  The Beam Condition Monitoring Leakage (BCML) system is a beam monitoring device in the CMS experiment at the LHC consisting of 32 poly-crystalline (pCVD) diamond sensors. The BCML sensors, located in rings around the beam, are exposed to high particle rates originating from the colli\-ding beams. These particles cause lattice defects, which act as traps for the ionized charge carrier leading to a reduced charge collection efficiency (CCE). The radiation induced CCE degradation was however much more severe than expected from low rate laboratory measurements. Measurement and simulations presented in this paper show that this discrepancy is related to the rate of incident particles. At high particle rates the trapping rate of the ionization is strongly increased compared to the detrapping rate leading to an increased build-up of space charge. This space charge locally reduces the internal electric field increasing the trapping rate and hence reducing the CCE even further. 
  }{%
 In order to connect these macroscopic measurements with the microscopic defects acting as traps for the ioni\-zation charge the TCAD simulation program \mbox{SILVACO} was used. It allows to introduce the defects as effective donor and acceptor levels and can calculate the electric field from Transient Current Technique (TCT) signals and CCE as function of the effective trap properties, like density, energy level and trapping cross section. After each irradiation step these properties were fitted to the data on the electric field from the TCT signals and CCE. Two effective acceptor and donor levels were needed to fit the data after each step. It turned out that the energy levels and cross sections could be kept constant and the trap density was proportional to the cumulative fluence of the irradiation steps. The highly non-linear rate dependent diamond polarization and the resulting signal loss can be simulated using this effective defect model and is in agreement with the measurement results.}}

%
%
%

\maketitle   

\begin{figure}[tbh]%
\includegraphics*[width=\linewidth]{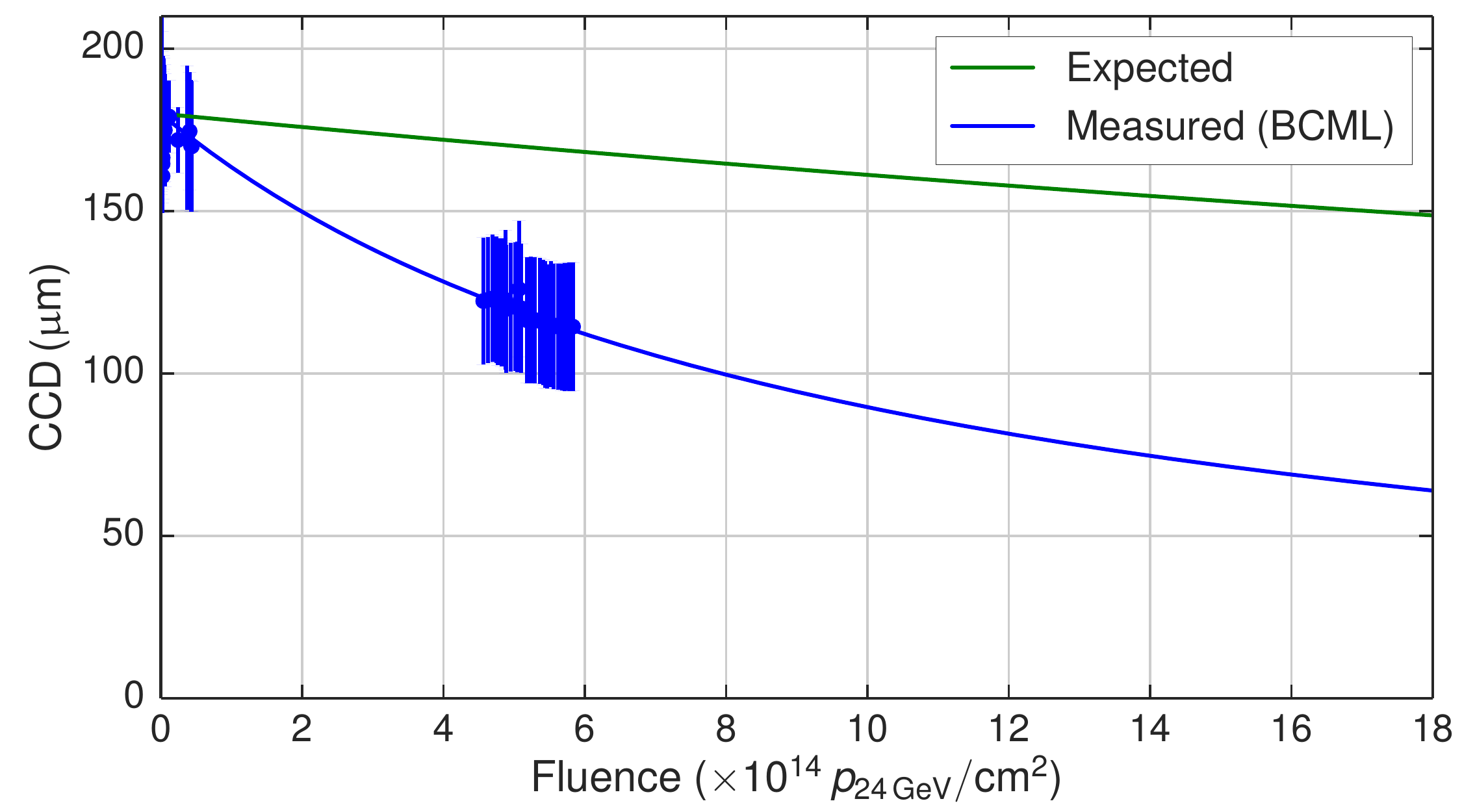}
\caption{%
  Comparison between expected (green) and measured (blue) CCD, or equivalently the radiation induced signal degradation of pCVD diamond sensors used at the CMS detector in the BCML system. The expected signal degradation is based on measurements done by the RD42 collaboration \cite{RD42_2008}. The measured curve shows a fit to the experimental data \cite{Kassel2017}.}
\label{BCML_signalloss}
\end{figure}

\section{Introduction}
The CMS Beam Condition Monitor Leakage (BCML) system at LHC is a beam monitoring device based on 32 poly-crystalline (pCVD) diamond sensors. The BCML sensors measure the ionization current created by beam losses leaking outside the beam pipe, e.g. by scattering on the residual gas or beam collimators. In case of very intense beam loss events, which could potentially damage the CMS detector, the BCML system triggers the LHC beam abort leading to a beam dump to protect the CMS detector.\\
Although diamond sensors were expected to be radiation hard, the charge collection efficiency (CCE) dropped much faster \cite{Guthoff2013168,Guthoff2015,Guthoff2014} in this high particle rate environment in comparison to low particle rate laboratory measurements \cite{RD42} and simulations \cite{deBoer2007,Guthoff2013}, see Fig.~\ref{BCML_signalloss}. Here the charge collection distance (CCD) of the BCML sensors with an average thickness of $\rm{d}=400\,\mu\rm{m}$ is plotted as funtion of particle fluence. The CCD defines the average drift length of the charge carriers before beeing trapped. After a total fluence of $\Phi = 16\times 10^{14}\,p_{\rm{24GeV}}/\rm{cm}^2$ the CCD of the pCVD diamonds used at the CMS detector dropped about twice as fast compared to the expectations based on laboratory measurements. This discrepancy in CCD between the real application in a particle detector and laboratory experiments can be explained by the rate dependent polarization \cite{Kassel2016,Rebai2016,Valentin2015} of the diamond detector, as was deduced from detailed laboratory measurements and simulations. The rate dependent diamond polarization describes the asymmetrical build-up of space charge in the diamond bulk, which leads to a locally reduced electric field configuration. The charge carrier recombination is increased in this low field region resulting in a reduced CCE of the diamond detector. At high particle rates the trapping rate of the ionization is even more increased compared to the de-trapping rate leading to an increased build-up of space charge. The increased amount of space charge causes an even stronger local reduction in the internal electric field and hence reduces the CCE further. The study presented in the following is an update to the work presented in \cite{Kassel2016}.

\section{Effective defect model describing the radiation induced signal degradation of diamond detectors}
Within the scope of this publication an effective defect model will be introduced, capable of explaining the radiation induced signal degradation of diamond detectors. This effective defect model was found by optimizing simulations to experimental Transient-Current-Technique (TCT) \cite{Isberg2002,Pernegger2005} and CCE measurement results of an irradiation campaign with high-quality single-crystal CVD (sCVD) diamonds detectors. The basic properties of this effective defect model like energy level and charge carrier cross sections for electrons and holes are based on \cite{Kassel2016}.

\begin{figure*}
\captionsetup[subfigure]{justification=centering}
\subfloat[un-irradiated]{%
\includegraphics*[width=0.3\textwidth]{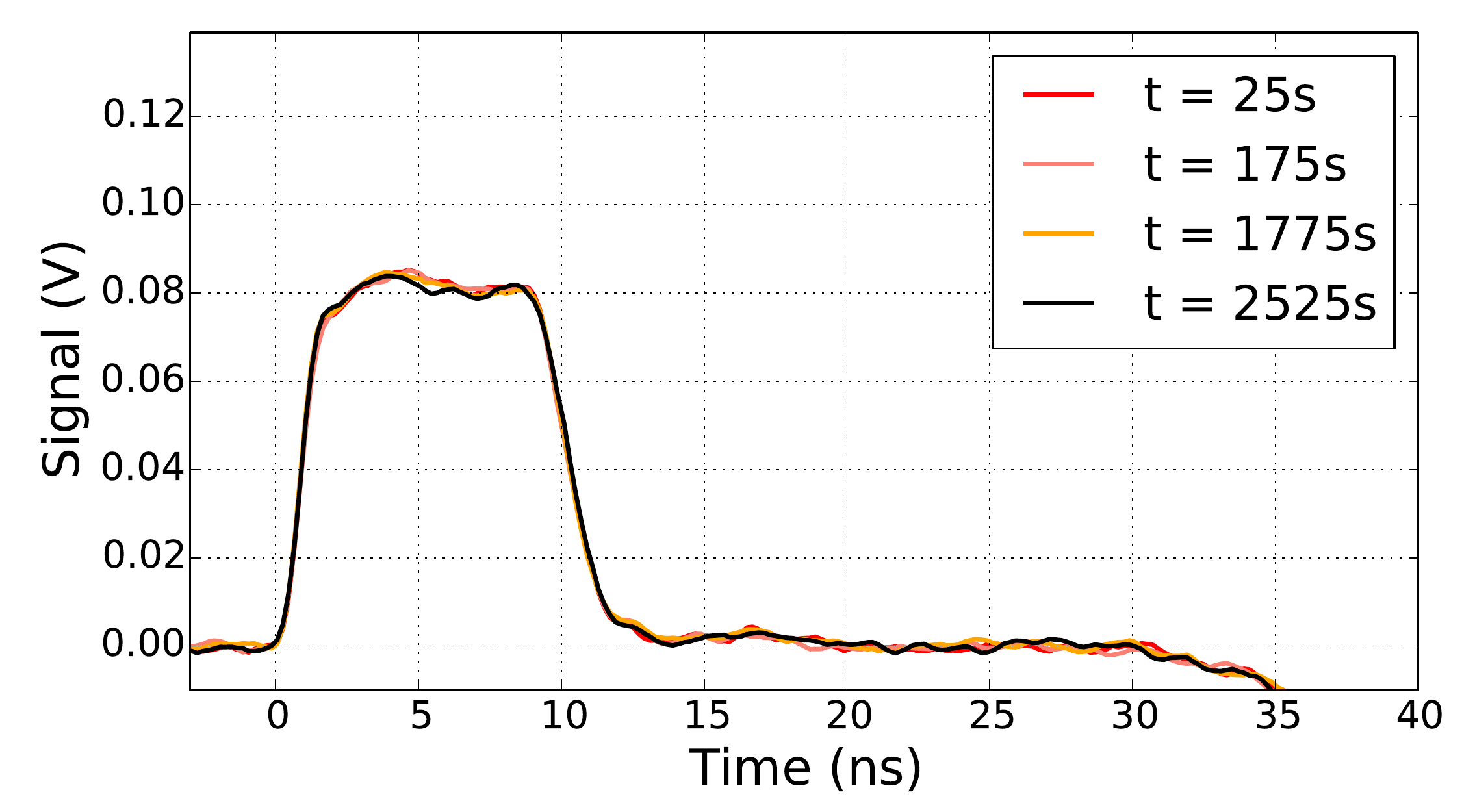}\label{Fig_Chapter6_TCTPol_200V_Hole_0}}\hfill
\subfloat[$\Phi=0.6\times 10^{13}\,n_{\rm{1\,MeV\,eq.}}/\rm{cm}^{2}$]{%
\includegraphics*[width=.3\textwidth]{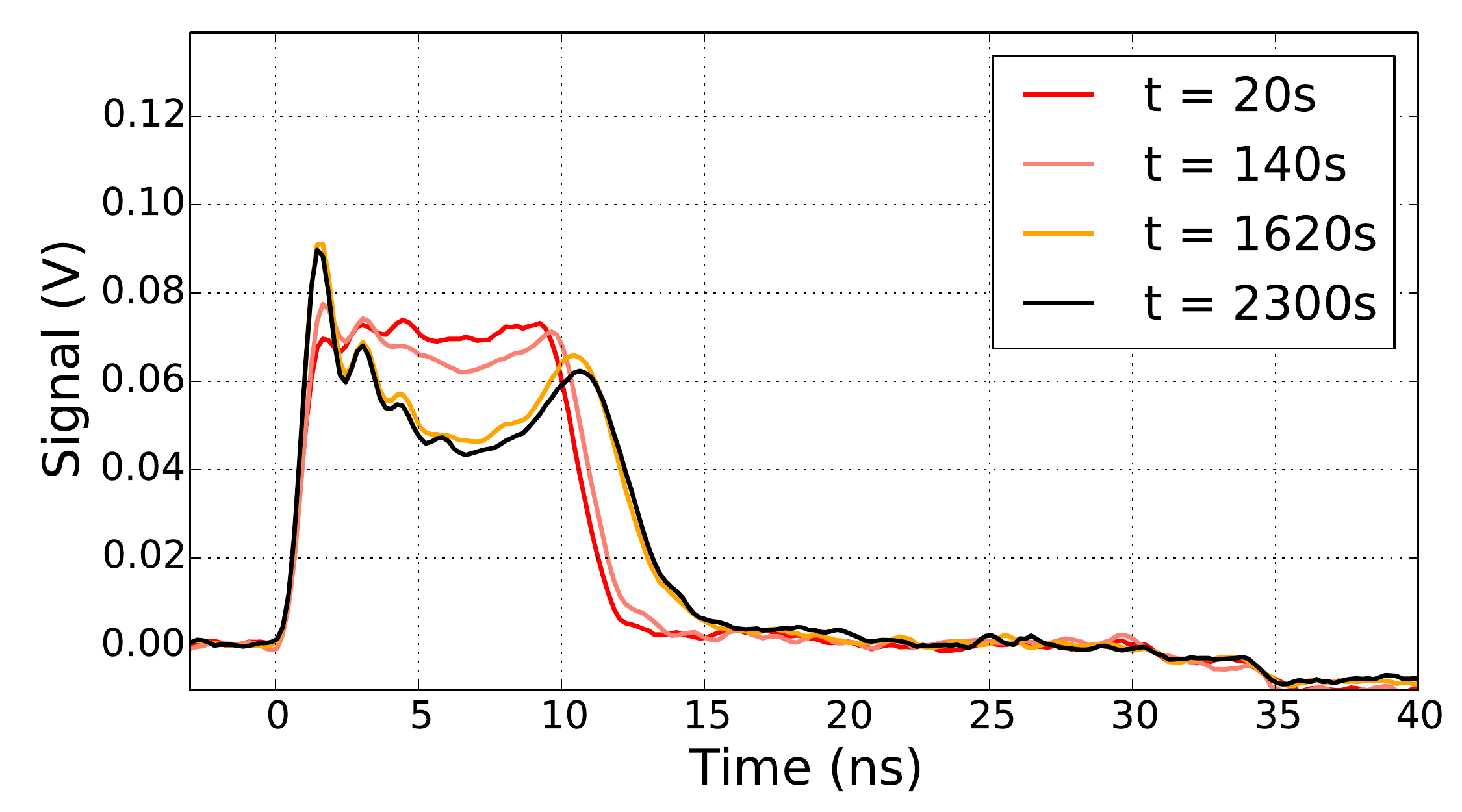}\label{Fig_Chapter6_TCTPol_200V_Hole_1}}\hfill
\subfloat[$\Phi=1.2\times 10^{13}\,n_{\rm{1\,MeV\,eq.}}/\rm{cm}^{2}$]{%
\includegraphics*[width=.3\textwidth]{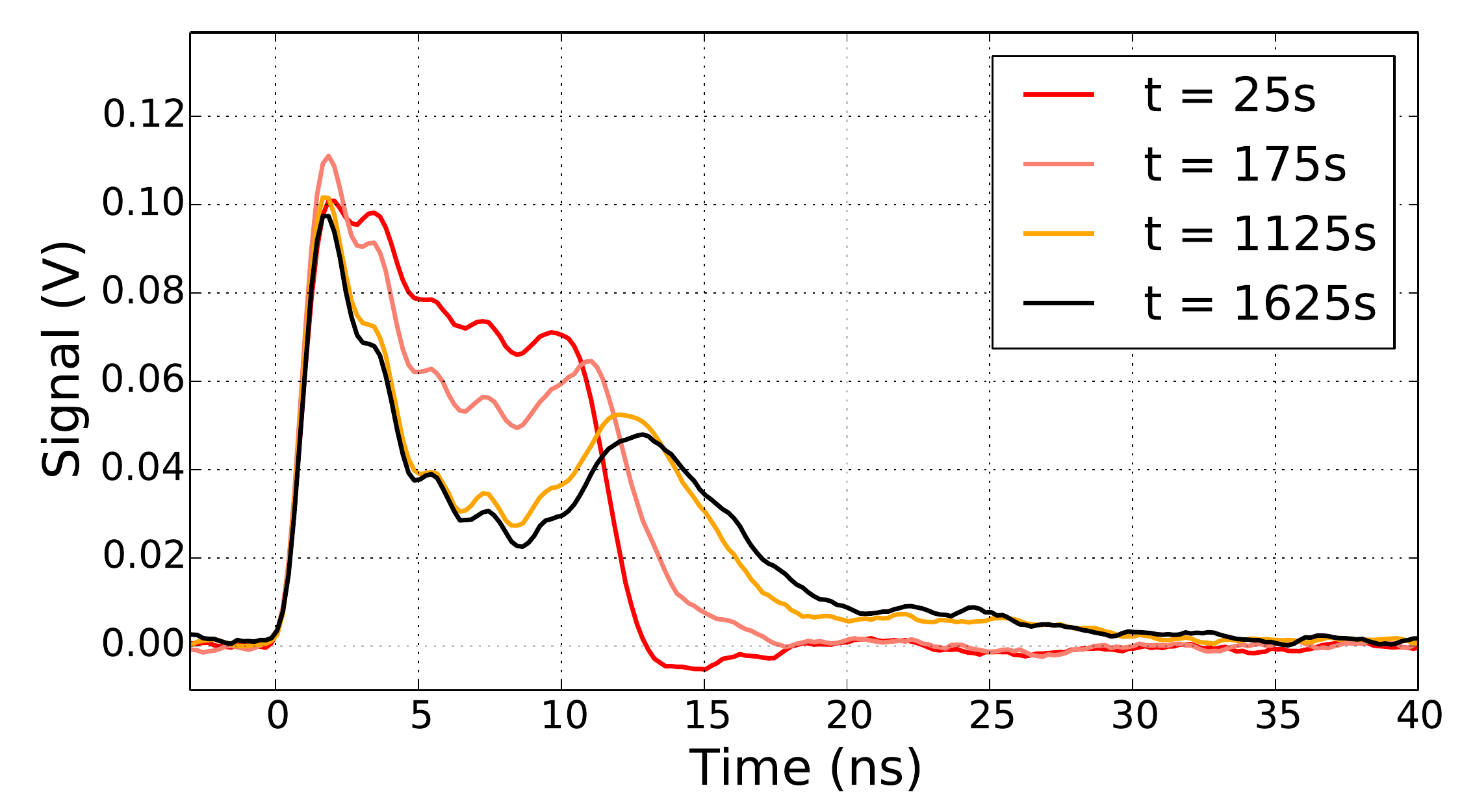}\label{Fig_Chapter6_TCTPol_200V_Hole_2}}\hfill
\subfloat[$\Phi=1.8\times 10^{13}\,n_{\rm{1\,MeV\,eq.}}/\rm{cm}^{2}$]{%
\includegraphics*[width=.3\textwidth]{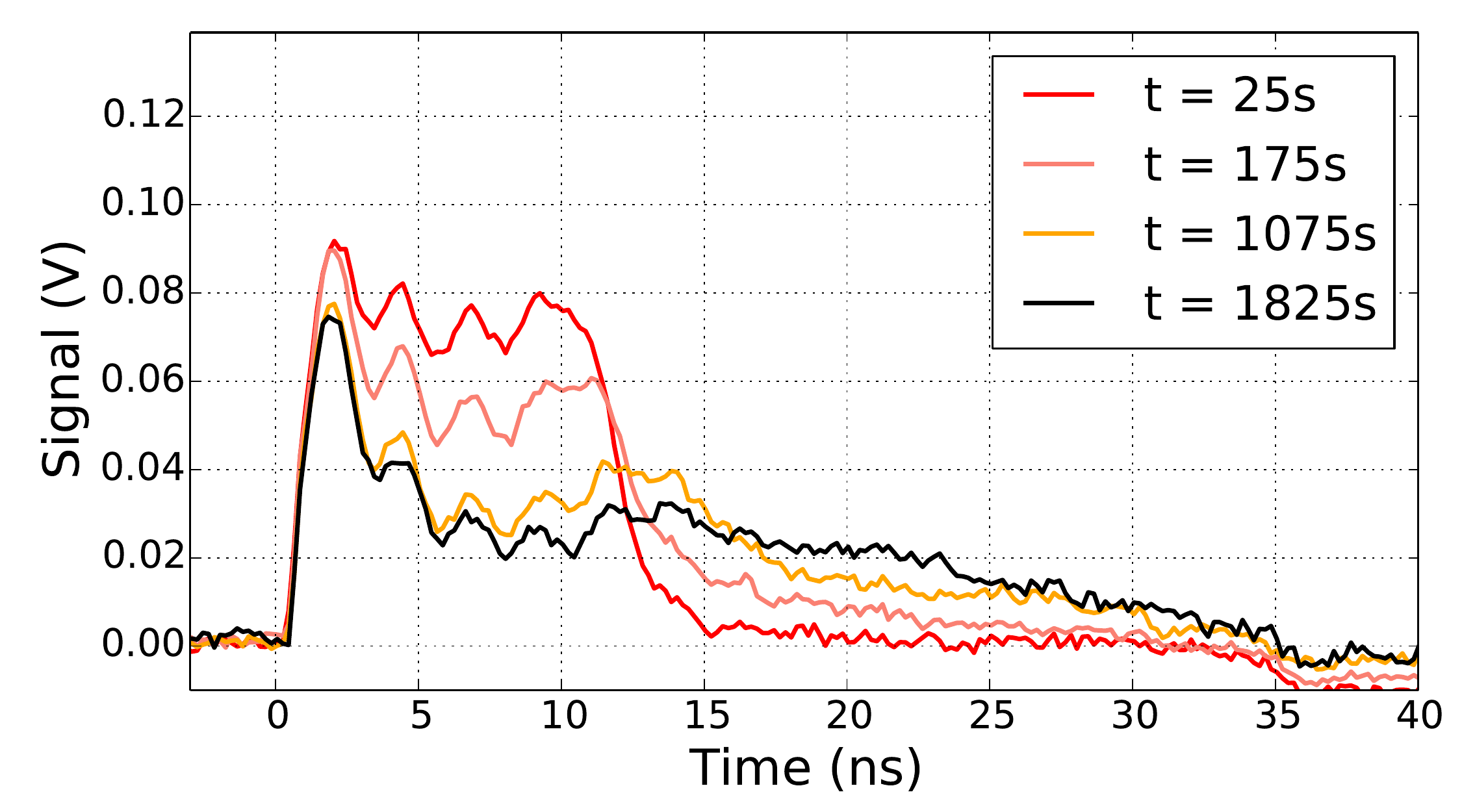}\label{Fig_Chapter6_TCTPol_200V_Hole_3}}\hfill
\subfloat[$\Phi=4.8\times 10^{13}\,n_{\rm{1\,MeV\,eq.}}/\rm{cm}^{2}$]{%
\includegraphics*[width=.3\textwidth]{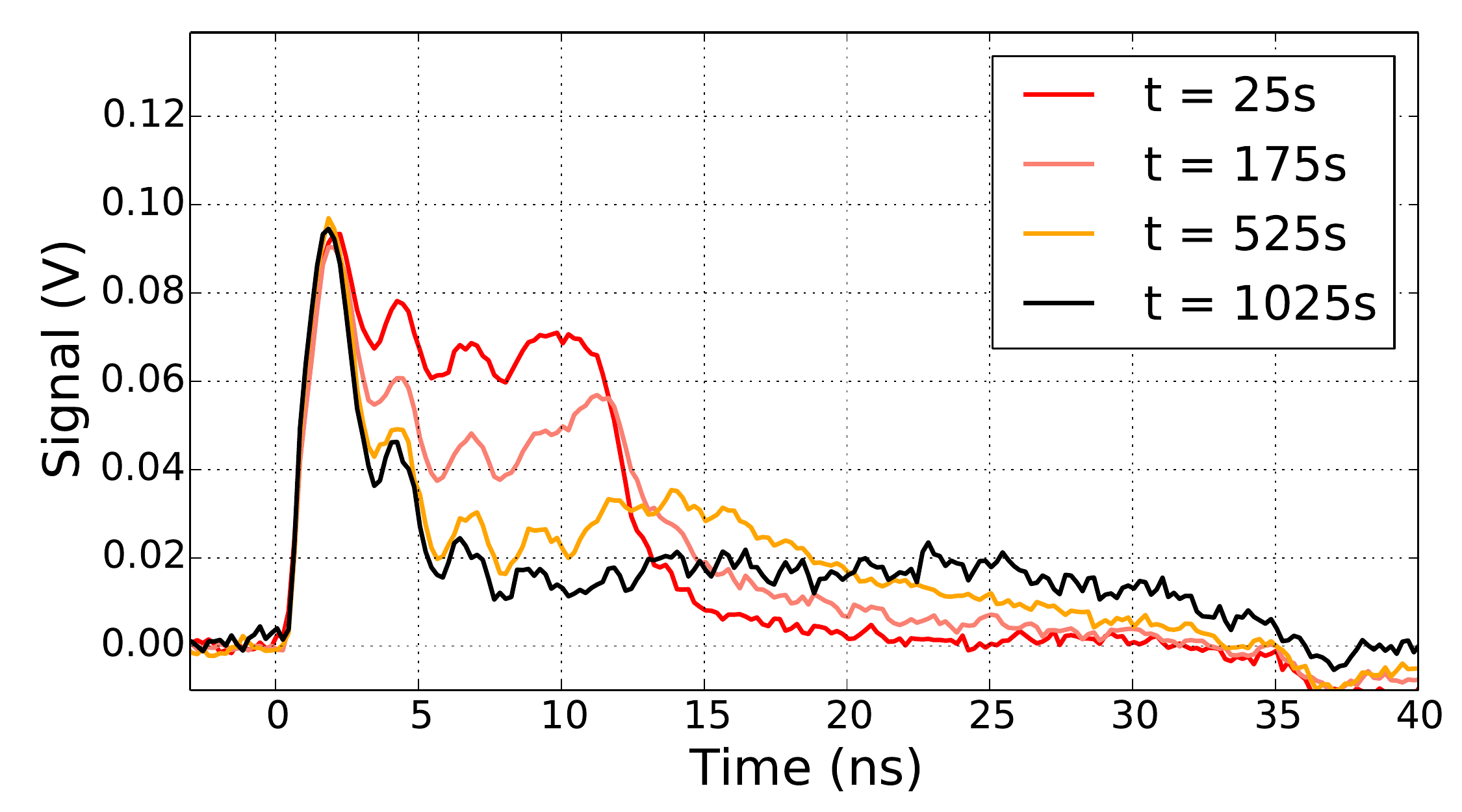}\label{Fig_Chapter6_TCTPol_200V_Hole_4}}\hfill
\subfloat[$\Phi=15.1\times 10^{13}\,n_{\rm{1\,MeV\,eq.}}/\rm{cm}^{2}$]{%
\includegraphics*[width=.3\textwidth]{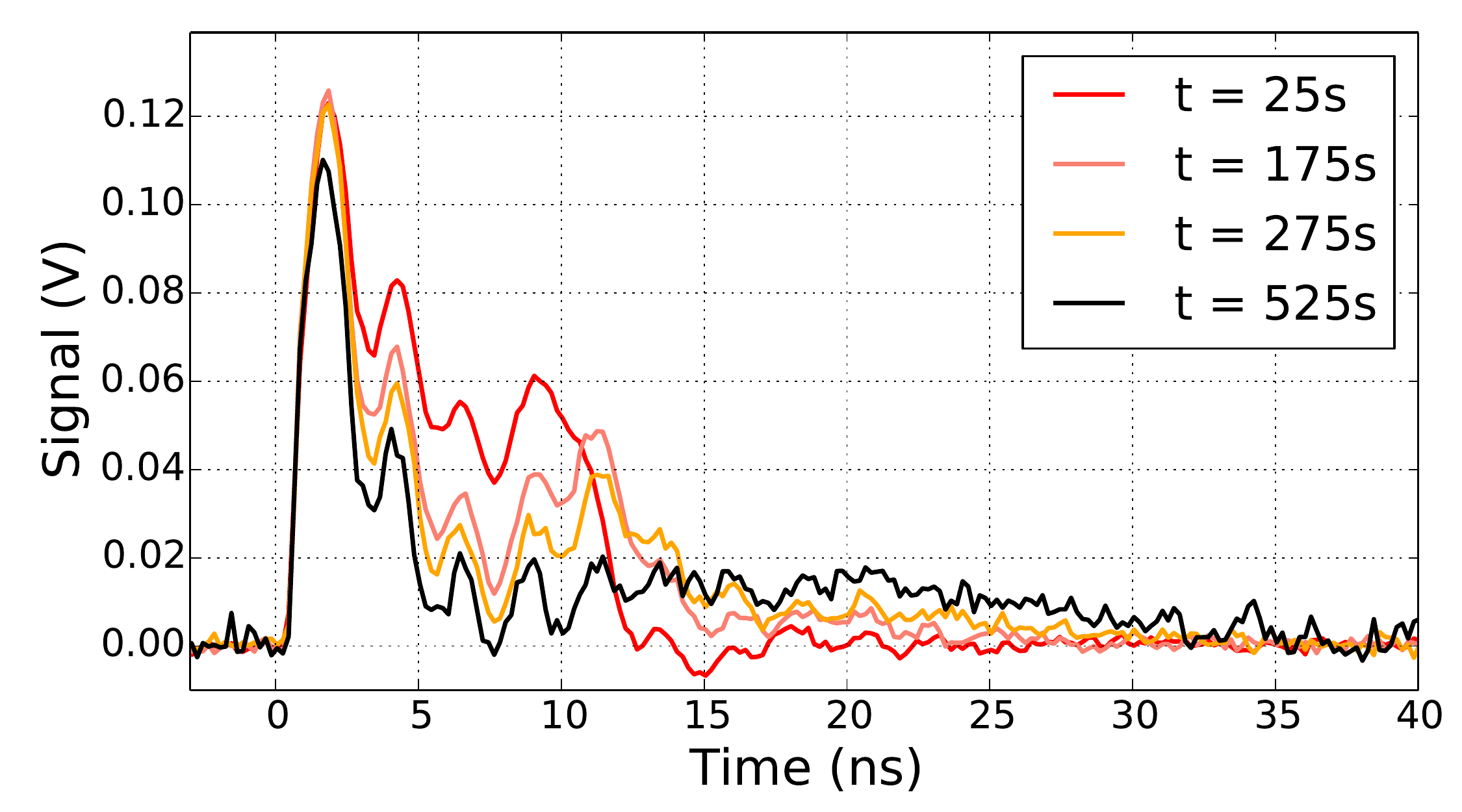}\label{Fig_Chapter6_TCTPol_200V_Hole_5}}\hfill
\caption{TCT measurements of the hole charge carriers at an electrical field of ${E = 0.36\,\rm{V}/\mu\rm{m}}$ for different irradiation damages. The increased radiation damage causes an increased build-up of space charge as function of exposure time to the $^{90}Sr$ source.}
\label{Fig_Chapter6_TCTPol_200V_Hole}
\end{figure*}

\subsection{Diamond irradiation campaign}
A dedicated irradiation campaign was carried out to gain quantitative understanding of the diamond polarization affecting the charge collection efficiency of diamond detectors. The diamond samples were irradiated stepwise up to a maximum fluence of $\Phi = 30.1\times 10^{13}\,\,n_{\rm{1\,MeV,eq.}}/\rm{cm}^2$. After each irradiation step the electrical properties of the diamond sensors were studied using the TCT method for an indirect electric field measurement and by measuring the charge collection efficiency. The diamond polarization, crucial in understanding the irradiation damage, is modifying the internal electrical field, which can be measured using TCT. The CCE measurements were used to study the reduced detector efficiency due to these electrical field modifications. 

In order to measure the build-up of polarization in the diamond for a given radiation damage the following measurement procedure is used for the TCT and CCE measurements:
\begin{enumerate}
	\item The diamond is exposed during the entire measurement to a constant ionization rate by a $^{90}Sr$ source creating electron-hole pairs in the entire diamond bulk filling up the traps. In the steady-state the trapping and detrapping rates are in equilibrium. According to the simulations discussed below about 55\,\% of the effective deep traps are filled.
	\item In order to remove any residual field and set the diamond into a unpolarized state, the sensor is exposed to the $^{90}Sr$ source for a duration of 20\,minutes without bias voltage applied. A homogeneous trap filling in the diamond bulk, and hence an unpolarized diamond state, is reached.
	\item The bias voltage is ramped up fast ($t_{\rm{ramp}}\leq 10\,\rm{s}$) and the measurement is started immediately ($t=0\,\rm{s}$).
	\item The diamond starts to polarize as soon as bias voltage is applied. The measurement is performed over an extended period of time ($t>3000\,\rm{s}$) until the diamond is fully polarized and the measurement results are stable.
\end{enumerate}

Four new single crystalline diamonds of highest quality 'electronic grade' corresponding to small nitrogen and boron impurities ($\left[N\right] < 5\,\rm{ppB}$ and $\left[B\right] < 1\,\rm{ppB}$), produced by Element6 \cite{Element6} were used to investigate the radiation induced signal degradation. The diamond samples were irradiated stepwise with either $23\,\rm{MeV}$ protons or with neutron particles with an energy distribution up to $10\,\rm{MeV}$ \cite{NeutronFacility}. More detailed information to the diamond samples, the TCT and CCE measurement setup can be found in \cite{Kassel2016}.

\paragraph{TCT measurement results}
The TCT measurements of the different irradiated diamond samples are shown in Fig.~\ref{Fig_Chapter6_TCTPol_200V_Hole} for the hole carrier drift. The expected rectangular TCT pulse shape, indicating a constant electrical field, is measured for the un-irradiated diamond sensor and remains stable as function of exposure time to the $^{90}Sr$ source. This rectangular TCT pulse shape is measured as well for the irradiated diamond samples immediately after the TCT measurement has been started ($t=25\,\rm{s}$) even though radiation damage leads to an increased amount of defects trapping free charge carriers. In this initial moment the diamond is in a pumped state and the defects are homogeneously ionized leading to a neutral effective space charge. Applying of bias voltage changes however the charge carrier distribution resulting in an inhomogeneous trapping. Trapped charge carriers, acting as space charge, are modifying the electric field configuration and leading to a modified TCT pulse shape. Increased radiation damage leads to an even stronger TCT pulse mo\-dification as function of the ionization duration. Furthermore, a faster transition to the final stable TCT pulse and hence to the final electrical field configuration is measured with respect to increased radiation damage. These TCT measurements demonstrate the increased build-up of space charge which leads to a strongly modified electric field distribution caused by radiation damage. The TCT measurement results for the electron drift are in agreement with the hole drift measurements and are therefore not explicitly discussed here. A direct comparison of the electron and hole drift can be found in \cite{Kassel2016}.

\begin{figure}%
\includegraphics*[width=\linewidth]{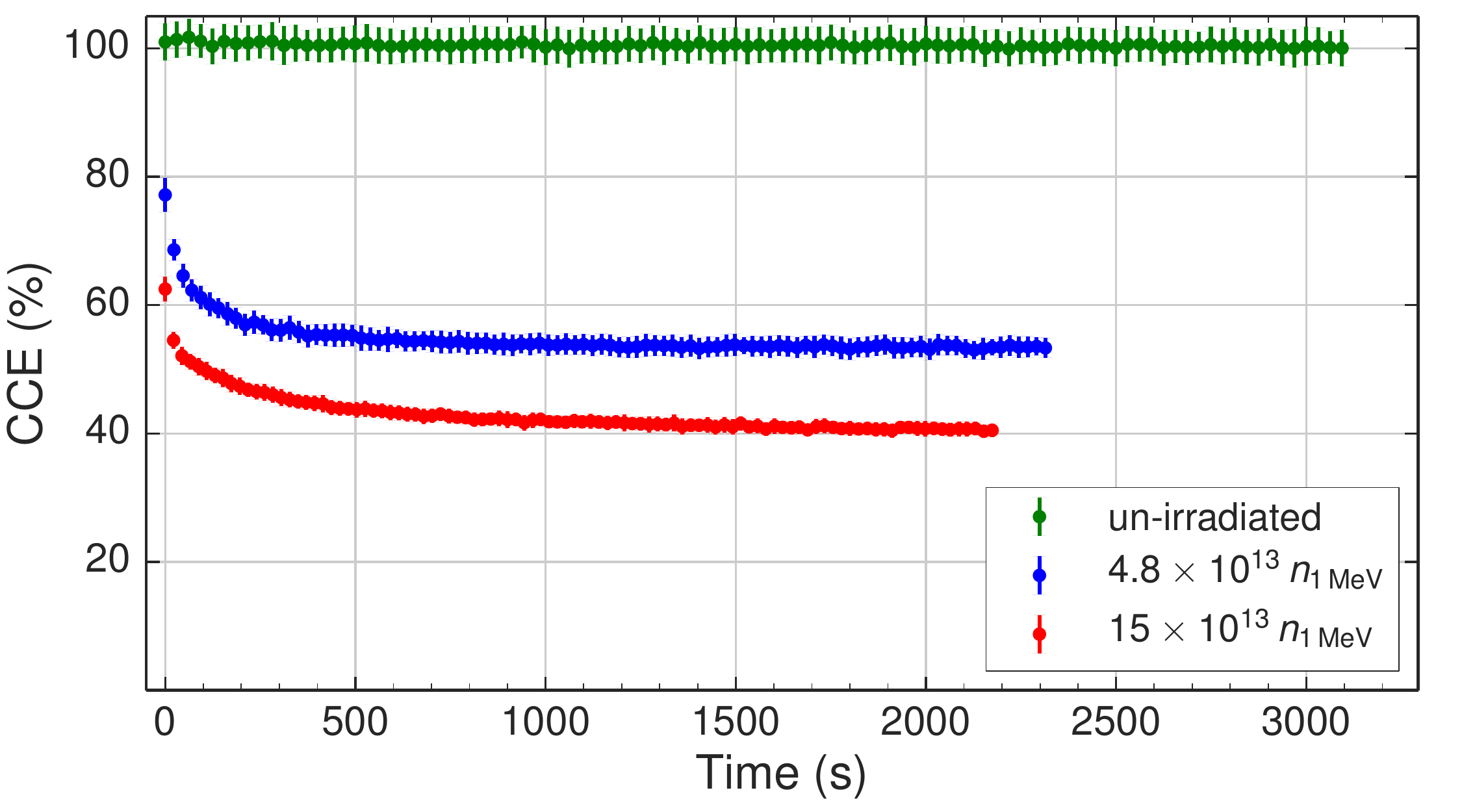}
\caption{%
  CCE measurement as function of exposure time to the ionization source for different irradiated diamond samples. The CCE measurements were done at an electric field of ${E = 0.36\,\rm{V}/\mu\rm{m}}$.}
\label{CCE_200V_1}
\end{figure}

\paragraph{CCE measurement results}\label{CCE_chapter}
The charge collection efficiency of the diamond sensors used in the irradiation campaign was measured regularly at the CCE setup at DESY in Zeuthen \cite{Grah2009} using minimum ionizing particles (MIP). The CCE measurement results are discussed in the following as function of measurement time, during which the diamond is exposed to an ionization source ($^{90}Sr$). The influence of the ionization source to the CCE measurement result is shown in Fig.~\ref{CCE_200V_1} for different radiation damages. The un-irradiated diamond sample is not influenced by the ionization source and remains constant at a CCE of 100\%. The radiation damaged diamond sensors are however affected by the ionization source. A steep decrease of the initial measured CCE is observed. The time constant of this reduction matches the time constant of the TCT pulse modification. Hence the modified electric field distribution directly affects the CCE and demonstrates the importance of the build-up of space charge in diamond sensors in order to understand the effects of radiation damage. This direct correlation of the TCT pulse modification and the measured CCE value is discussed in more detail in \cite{Kassel2016}.

\begin{figure}%
\includegraphics*[width=\linewidth]{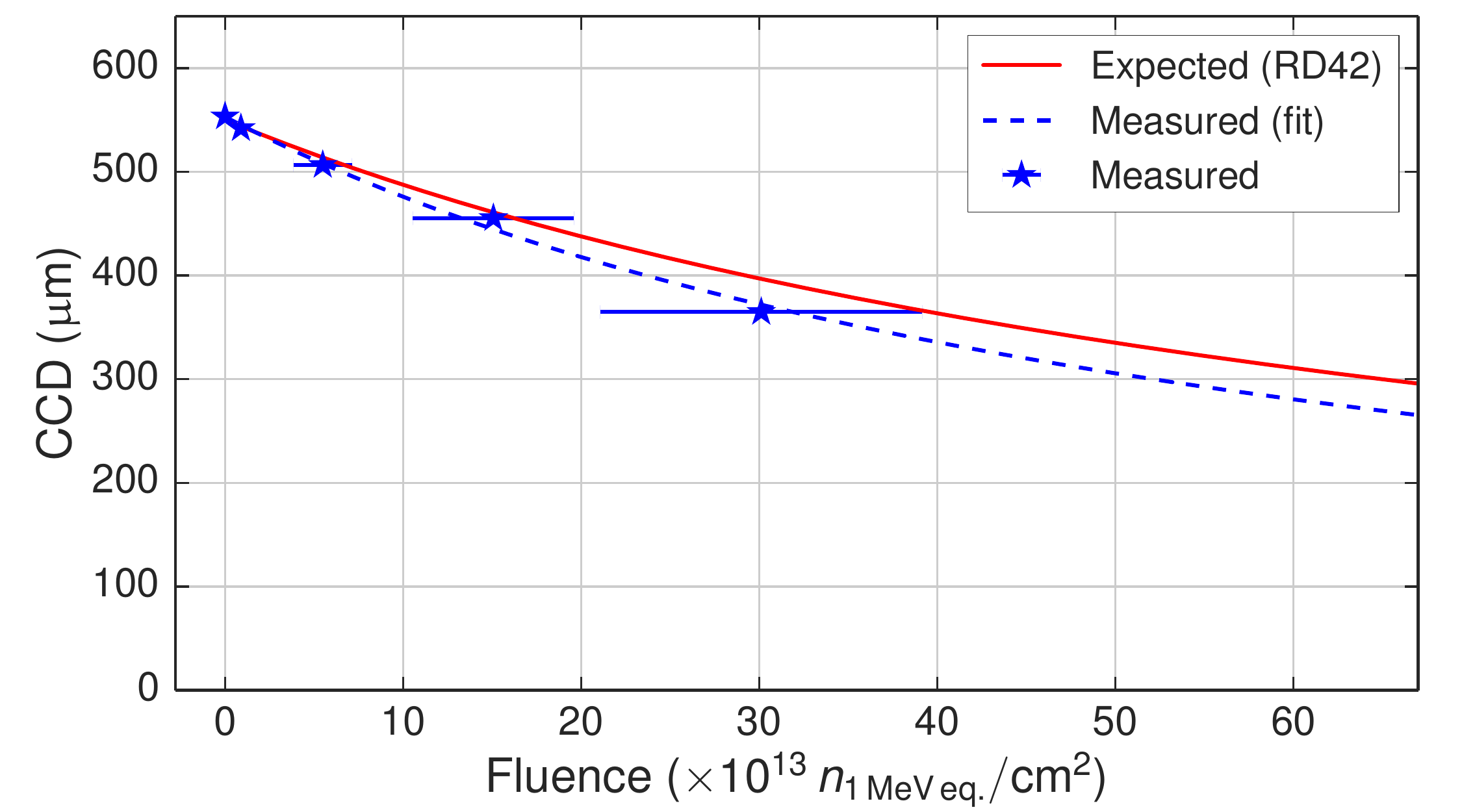}
\caption{%
  The CCD as function of fluence for an electrical field of ${E = 1.0\,\rm{V}/\mu\rm{m}}$. The results are compared to the expected radiation damage measured by the RD42 collaboration \cite{RD42_2008}.}
\label{CCE_200V_2}
\end{figure}

The stabilized CCD measurements are plotted in Fig.~\ref{CCE_200V_2} as function of fluence for an electrical field of ${E = 1.0\,\rm{V}/\mu\rm{m}}$. The CCD as function of radiation damage can be described by the following equation:
\begin{equation}
\frac{1}{\rm{CCD}(\Phi)} = \frac{1}{\rm{CCD}_0} + k \times \Phi,
\label{Eq_RD42_fin}
\end{equation}
with $\rm{CCD}_0$ as initial CCD, $\Phi$ as particle fluence in $p_{\rm{24GeV\,eq.}}/\rm{cm}^{2}$ and $k$ as radiation constant. It was found that the radiation constant $k$ describes both, the behavior of sCVD as well as pCVD diamond sensors \cite{RD42_2008}. Irradiation studies done by the RD42 collaboration with pCVD diamond sensors determined a radiation constant of $k = 6.5\times 10^{-19}\,\rm{cm}^2/\mu\rm{m}$ \cite{RD42_2008} for irradiation with 24\,GeV protons. The conversion of the radiation damage created by $n_{\rm{1\,MeV}}$ to $p_{\rm{24\,GeV}}$ is caluclated using NIEL \cite{Guthoff2014223}, which is increased by a factor of $3.59$.

Based on the measurements done within this paper a radiation constant of $k = (8.2 \pm 0.5)\times 10^{-19}\,\rm{cm}^2/\mu\rm{m}$ was found which is in reasonable agreement with the RD42 value.

\subsection{TCAD simulation of radiation damage}
In order to create a defect model for diamond sensors and gain quantitative understanding of the radiation induced signal degradation, the diamond sensor was modeled with the software Silvaco TCAD \cite{Silvaco}. Besides the electrical diamond properties, like e.g. band gap or mobility parameters, radiation induced lattice defects can be taken into account by introducing effective deep traps acting as recombination centers. The properties of these defects, like energy levels, capture cross section for electrons and holes, were found by optimizing the simulation of TCT and MIP pulses to match the experimental data and are listed in Table \ref{Table_Chapter6_Trap_OneDamage}, further information can be found in \cite{Kassel2016}. The simulated MIP pulses were used to calculate the charge collection efficiency of the diamond sensor. The TCAD simulation included furthermore a detailed implementation of the geometrical properties of the TCT and CCE measurement setups, affecting e.g. the energy deposition of the ionization sources in the diamond sensors.

\begin{table}[t]%
\centering
  \caption{Physical parameters of the effective recombination centers \cite{Kassel2016}. Both effective recombination centers $eRC1/2$ are present as acceptor- and donor-like traps.}
  \begin{tabular}[htbp]{@{}lcc@{}}
    \hline
    Trap name & $eRC1$ & $eRC2$ \\
    \hline
    Energy level $E_t$ (eV)  &  1.8 & 0.83 \\
    Cross section $\sigma_e$ ($\times 10^{-14}\,\rm{cm}^2$) & 0.7	& 2.0 \\
    Cross section $\sigma_h$ ($\times 10^{-14}\,\rm{cm}^2$) & 0.7	& 1.0 \\
    \hline
  \end{tabular}
  \label{Table_Chapter6_Trap_OneDamage}
\end{table}

\begin{figure*}
\captionsetup[subfigure]{justification=centering}
\subfloat[$\Phi=0.6\times 10^{13}\,n_{\rm{1\,MeV\,eq.}}\rm{cm}^{-2}$ $E = 0.18\,\rm{V}/\mu\rm{m}$][$\Phi=0.6\times 10^{13}\,n_{\rm{1\,MeV\,eq.}}\rm{cm}^{-2}$\\$E = 0.18\,\rm{V}/\mu\rm{m}$]{%
\includegraphics*[width=0.3\textwidth]{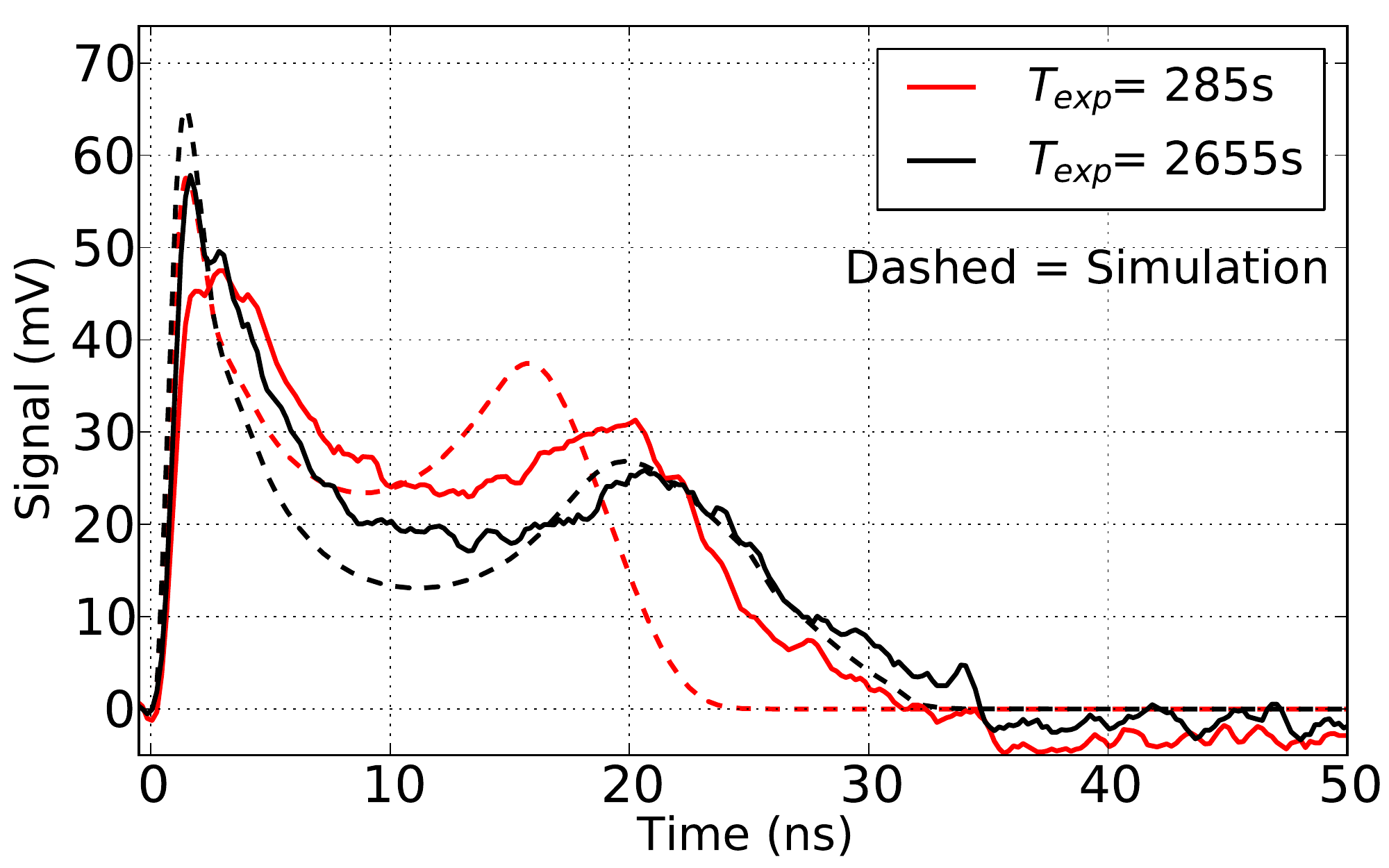}\label{Fig_Chapter6_TCTPol_200V_Hole_0}}\hfill
\subfloat[$\Phi=0.9\times 10^{13}\,n_{\rm{1\,MeV\,eq.}}\rm{cm}^{-2}$ $E = 0.18\,\rm{V}/\mu\rm{m}$][$\Phi=0.9\times 10^{13}\,n_{\rm{1\,MeV\,eq.}}\rm{cm}^{-2}$\\$E = 0.18\,\rm{V}/\mu\rm{m}$]{%
\includegraphics*[width=.3\textwidth]{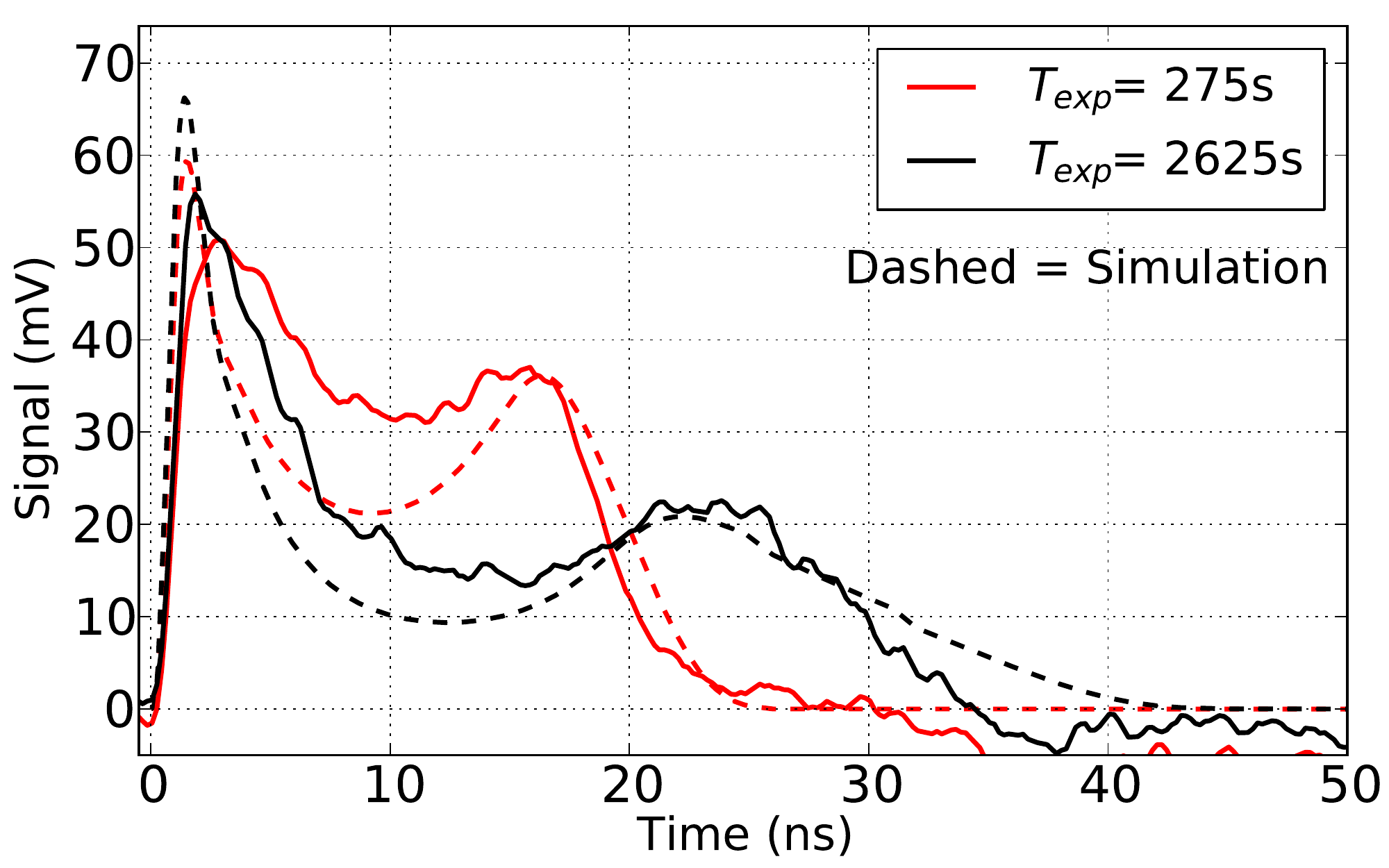}\label{Fig_Chapter6_TCTPol_200V_Hole_1}}\hfill
\subfloat[$\Phi=1.2\times 10^{13}\,n_{\rm{1\,MeV\,eq.}}\rm{cm}^{-2}$ $E = 0.36\,\rm{V}/\mu\rm{m}$][$\Phi=1.2\times 10^{13}\,n_{\rm{1\,MeV\,eq.}}\rm{cm}^{-2}$\\$E = 0.36\,\rm{V}/\mu\rm{m}$]{%
\includegraphics*[width=.3\textwidth]{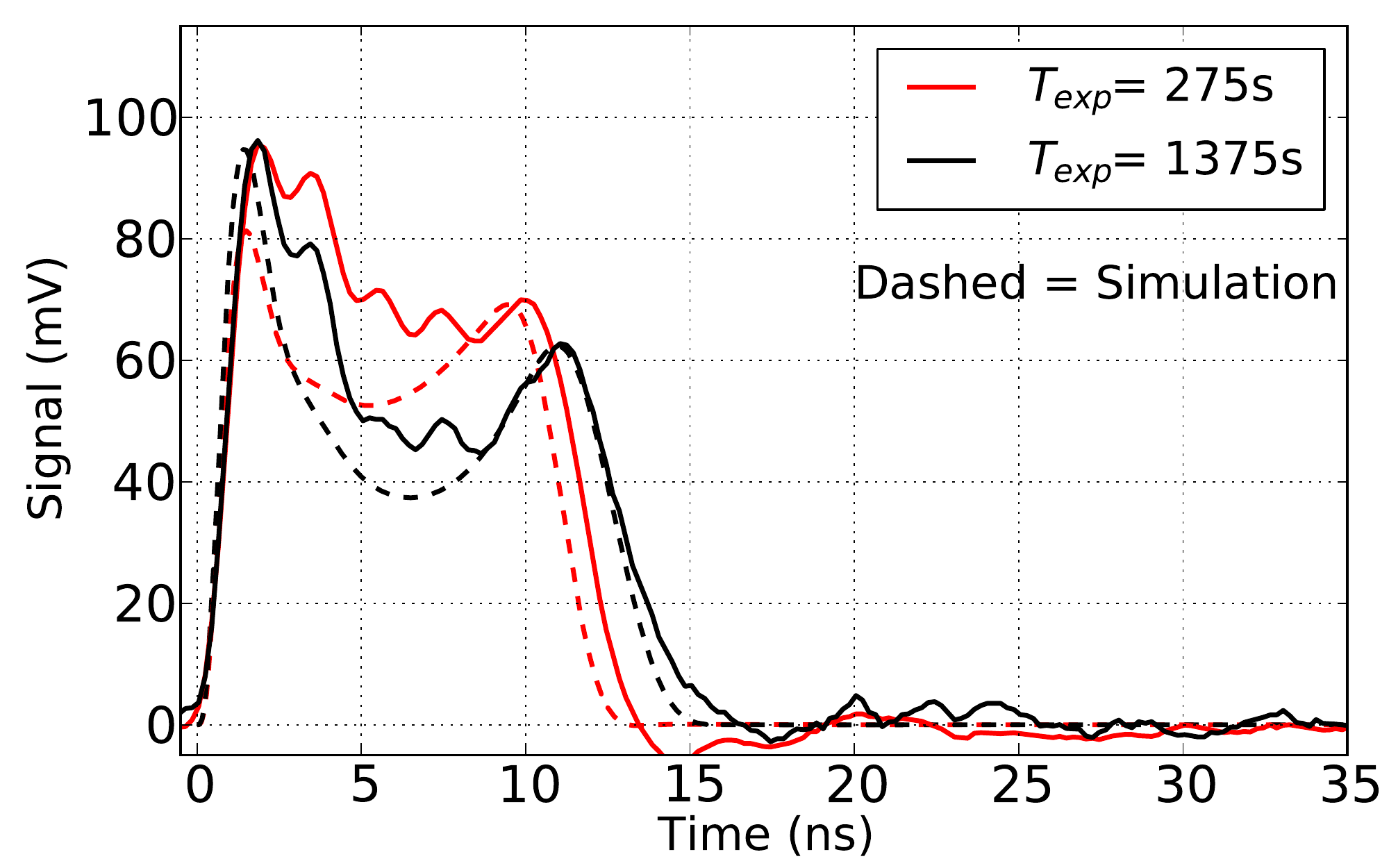}\label{Fig_Chapter6_TCTPol_200V_Hole_2}}\hfill
\subfloat[$\Phi=5.5\times 10^{13}\,n_{\rm{1\,MeV\,eq.}}\rm{cm}^{-2}$ $E = 0.36\,\rm{V}/\mu\rm{m}$][$\Phi=5.5\times 10^{13}\,n_{\rm{1\,MeV\,eq.}}\rm{cm}^{-2}$\\$E = 0.36\,\rm{V}/\mu\rm{m}$]{%
\includegraphics*[width=.3\textwidth]{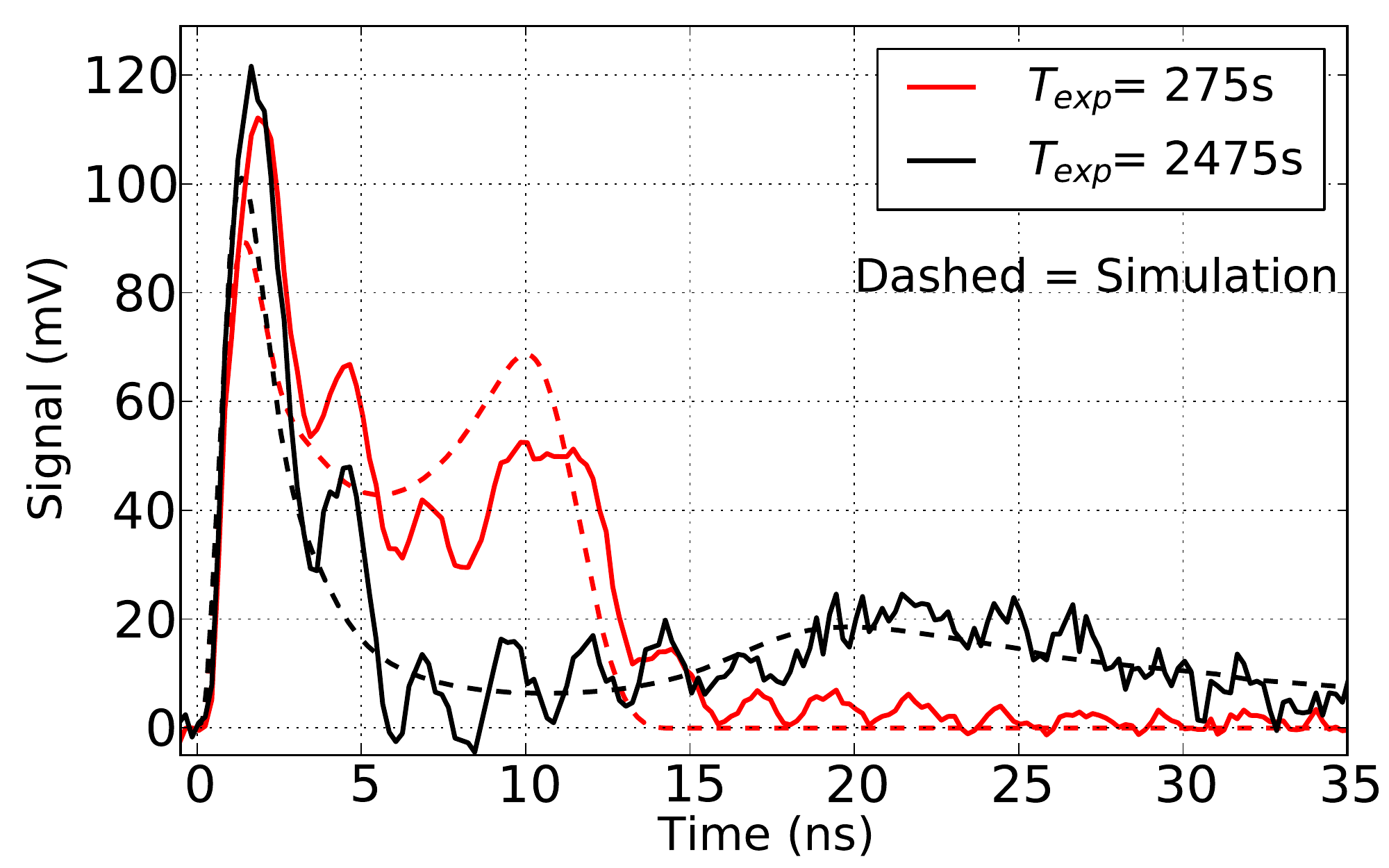}\label{Fig_Chapter6_TCTPol_200V_Hole_3}}\hfill
\subfloat[$\Phi=4.8\times 10^{13}\,n_{\rm{1\,MeV\,eq.}}\rm{cm}^{-2}$ $E = 0.73\,\rm{V}/\mu\rm{m}$][$\Phi=4.8\times 10^{13}\,n_{\rm{1\,MeV\,eq.}}\rm{cm}^{-2}$\\$E = 0.73\,\rm{V}/\mu\rm{m}$]{%
\includegraphics*[width=.3\textwidth]{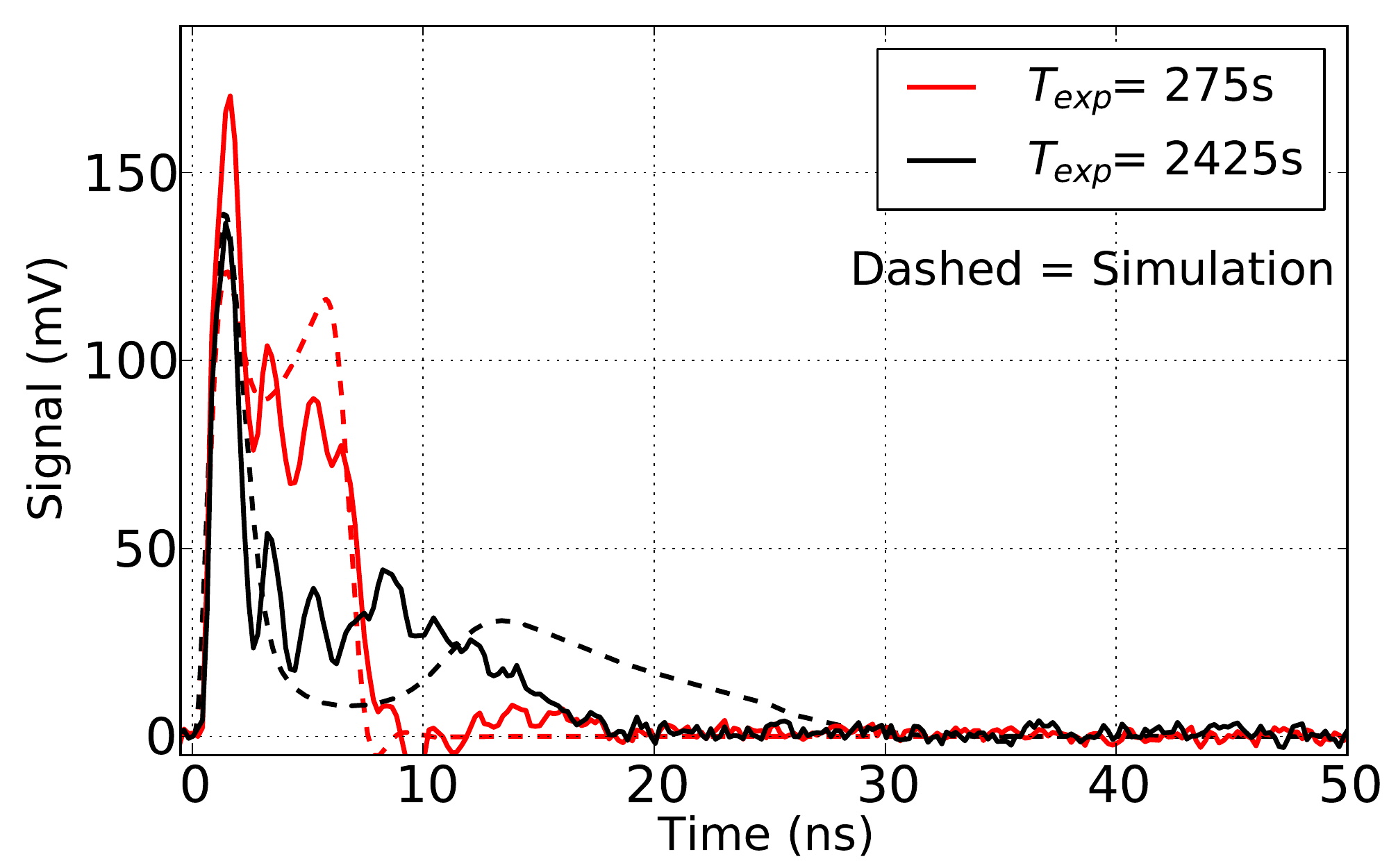}\label{Fig_Chapter6_TCTPol_200V_Hole_4}}\hfill
\subfloat[$\Phi=30.1\times 10^{13}\,n_{\rm{1\,MeV\,eq.}}\rm{cm}^{-2}$ $E = 1.09\,\rm{V}/\mu\rm{m}$][$\Phi=30.1\times 10^{13}\,n_{\rm{1\,MeV\,eq.}}\rm{cm}^{-2}$\\$E = 1.09\,\rm{V}/\mu\rm{m}$]{%
\includegraphics*[width=.3\textwidth]{07_2E14_27135474_TCT_RESULTS.pdf}\label{Fig_Chapter6_TCTPol_200V_Hole_5}}\hfill
\caption{Comparison between the TCT simulation and measurement for the hole charge carrier drift for different fluences varying from $\Phi=0.6\times 10^{13}\,n_{\rm{1\,MeV\,eq.}}\rm{cm}^{-2}$ to $\Phi=30.1\times 10^{13}\,n_{\rm{1\,MeV\,eq.}}\rm{cm}^{-2}$ in panel a) to f). The defect densities $\rho_{eRC1}$ and $\rho_{eRC2}$ were optimized for each fluence to match the measurement results.}
\label{Fig_Chapter6_TCTPol_200V_Hole_SIM}
\end{figure*}

\paragraph{Optimization of the effective defect densities to the TCT and CCE measurements}
Within the scope of the irradiation campaign TCT and CCE measurements for 12 different irradiation steps were obtained. The TCT and CCE simulation were fitted to each TCT and CCE measurement and a uniquely optimized trap configuration was found with respect to the particular fluence. The traps were optimized by adjusting the trap density of the effective recombination centers ($\rho_{eRC1}$ and $\rho_{eRC2}$), since the trap properties like energy level or cross section should not be affected by the fluence. In Fig.~\ref{Fig_Chapter6_TCTPol_200V_Hole_SIM} the simulated and measured TCT pulses of the hole charge carrier drift are shown for different fluences. The TCT simulation were optimized to the measurement results at the lowest measurable bias voltage, where the polarization is affecting the electric field most. The build-up of the internal polarization field is so quick for higher fluences that a reliable TCT measurement  at low bias voltages was not possible. The TCT measurements were therefore done at increased bias voltages.

\begin{figure*}
\captionsetup[subfigure]{justification=centering}
\subfloat[$\Phi=0.9\times 10^{13}\,n_{\rm{1\,MeV\,eq.}}\rm{cm}^{-2}$ $E = 0.18\,\rm{V}/\mu\rm{m}$][$\Phi=0.9\times 10^{13}\,n_{\rm{1\,MeV\,eq.}}\rm{cm}^{-2}$\\$E = 0.18\,\rm{V}/\mu\rm{m}$]{%
\includegraphics*[width=0.43\textwidth]{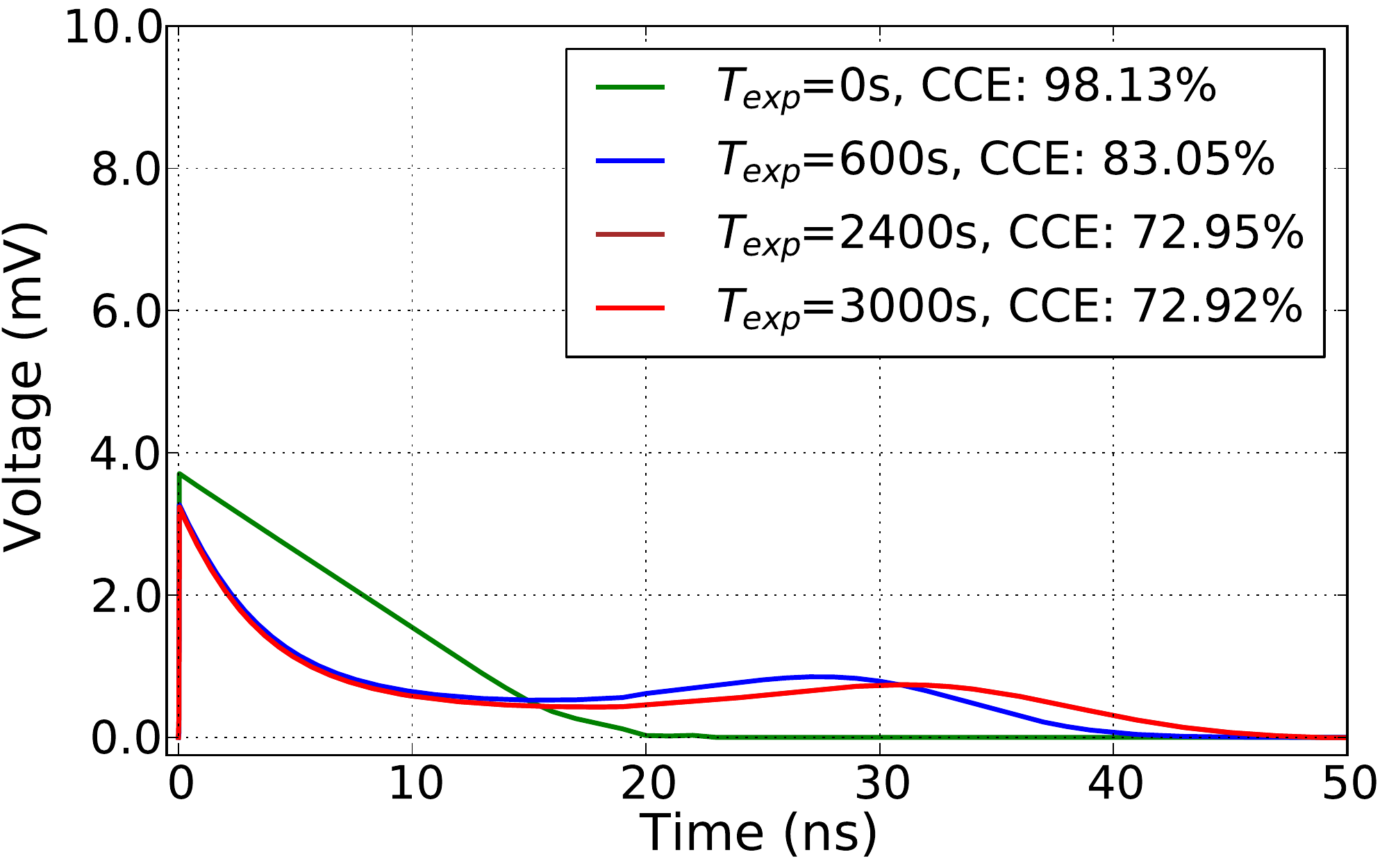}\label{Fig_EDM_MIP_1}}\hfill
\subfloat[$\Phi=30\times 10^{13}\,n_{\rm{1\,MeV\,eq.}}\rm{cm}^{-2}$ $E = 0.74\,\rm{V}/\mu\rm{m}$][$\Phi=30\times 10^{13}\,n_{\rm{1\,MeV\,eq.}}\rm{cm}^{-2}$\\$E = 0.74\,\rm{V}/\mu\rm{m}$]{%
\includegraphics*[width=.43\textwidth]{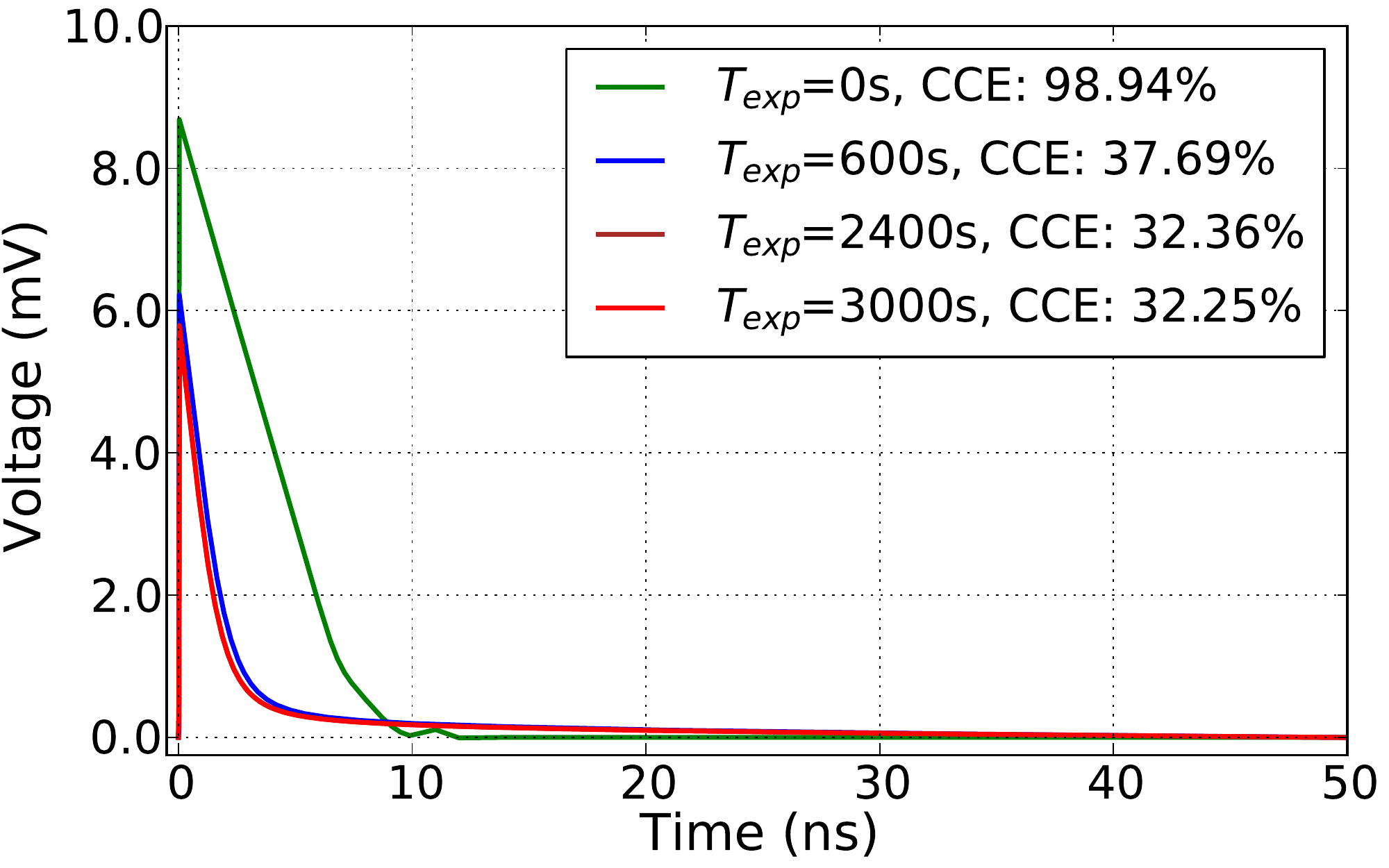}\label{Fig_EDM_MIP_2}}\hfill
\subfloat[$\Phi=0.9\times 10^{13}\,n_{\rm{1\,MeV\,eq.}}\rm{cm}^{-2}$ $E = 0.18\,\rm{V}/\mu\rm{m}$][$\Phi=0.9\times 10^{13}\,n_{\rm{1\,MeV\,eq.}}\rm{cm}^{-2}$\\$E = 0.18\,\rm{V}/\mu\rm{m}$]{%
\includegraphics*[width=0.47\textwidth]{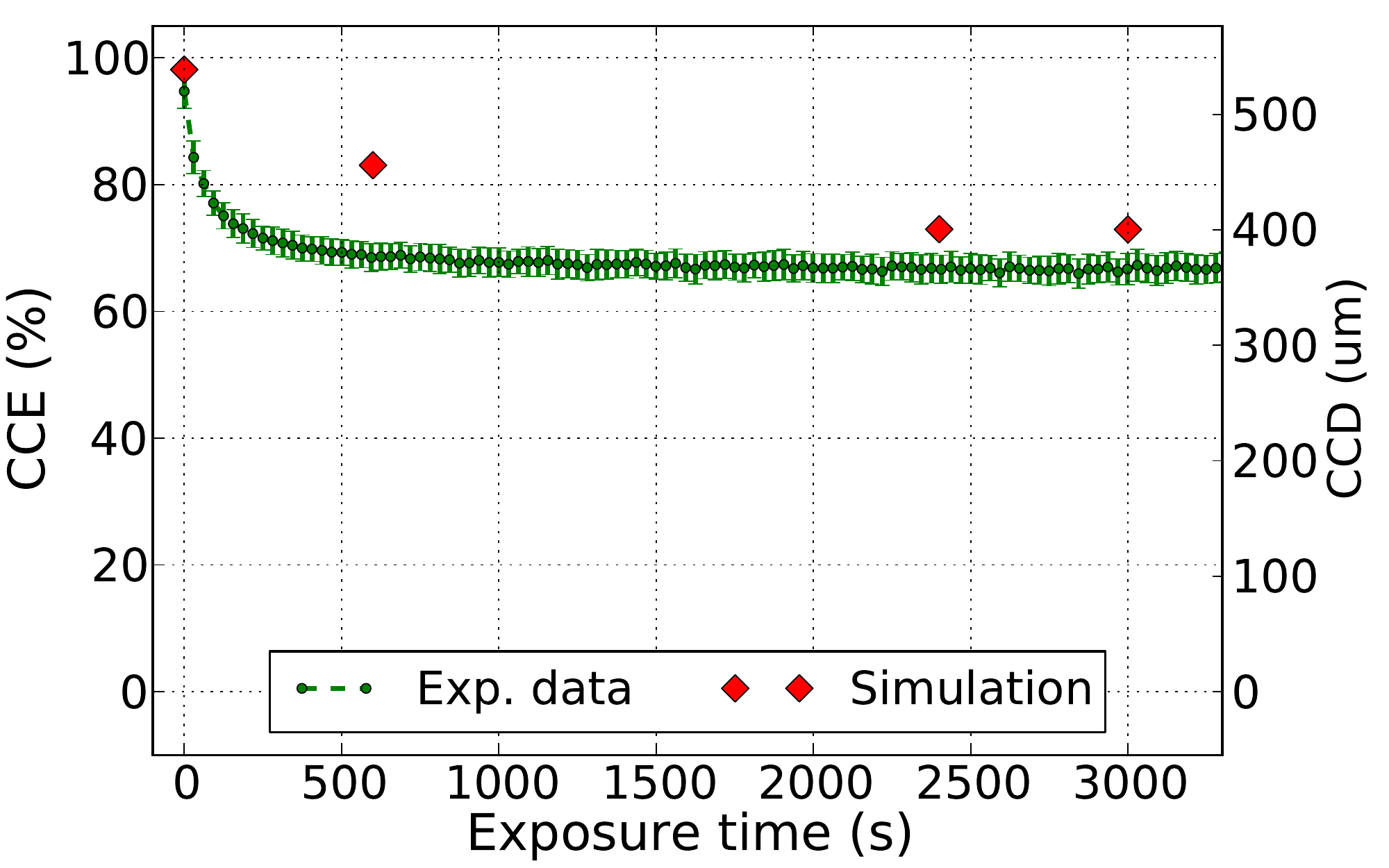}\label{Fig_EDM_CCE_1}}\hfill
\subfloat[$\Phi=30\times 10^{13}\,n_{\rm{1\,MeV\,eq.}}\rm{cm}^{-2}$ $E = 0.74\,\rm{V}/\mu\rm{m}$][$\Phi=30\times 10^{13}\,n_{\rm{1\,MeV\,eq.}}\rm{cm}^{-2}$\\$E = 0.74\,\rm{V}/\mu\rm{m}$]{%
\includegraphics*[width=.47\textwidth]{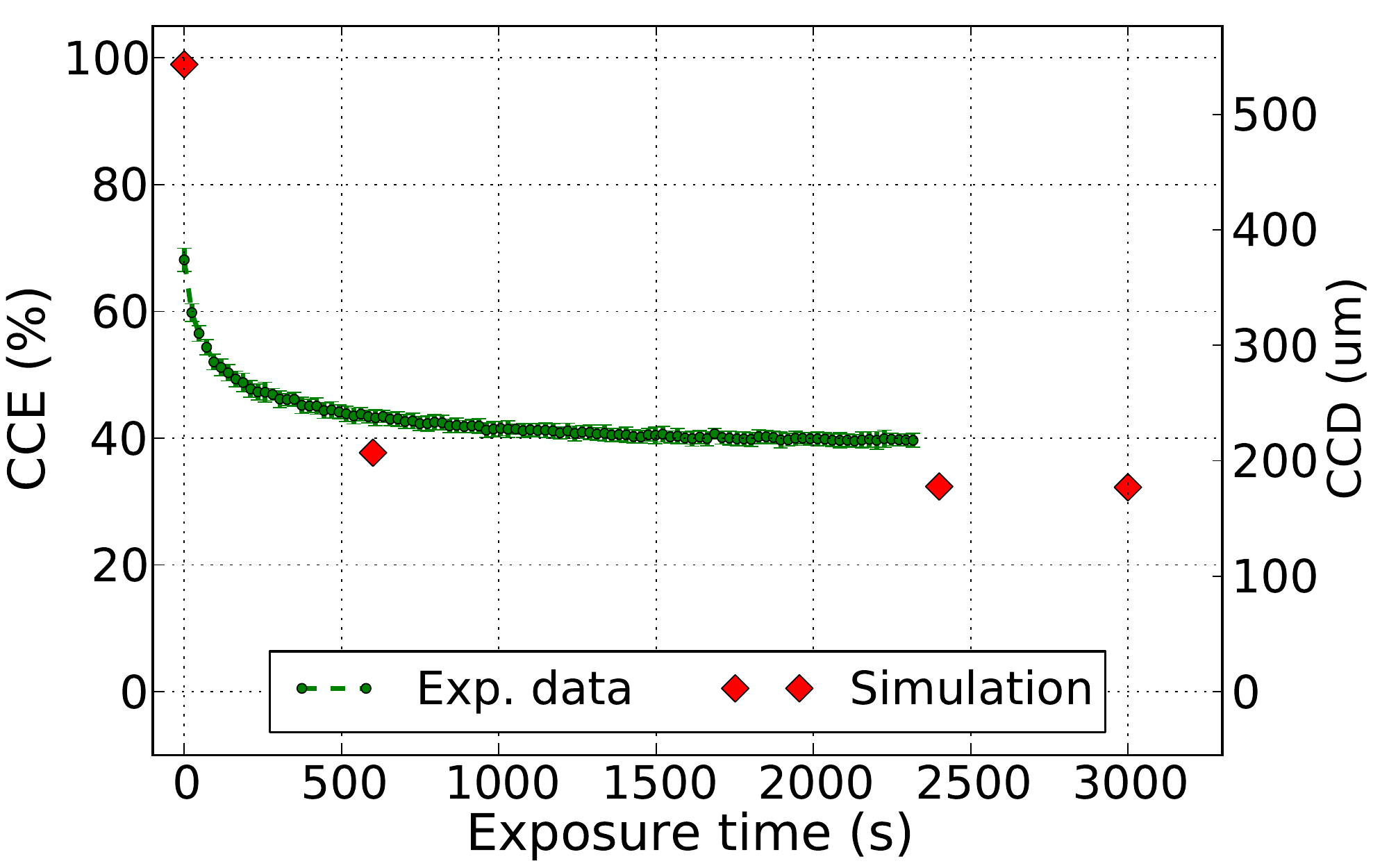}\label{Fig_EDM_CCE_2}}\hfill
\caption{Simulation of CCE as function of fluence for two different bias voltages. The top row shows the simulated MIP signal as function of polarization for two different fluences. The calculated CCE based on the integrated MIP signal is shown in the lower row as function of exposure time to the $^{90}Sr$ source.}
\label{Fig_EDM_MIP_CCE}
\end{figure*}

The simulated MIP signal is shown in Fig.~\ref{Fig_EDM_MIP_1},b as function of the ionization time ($T_{exp}$) for two different fluences. The MIP signal was integrated to calculate the charge collection efficiency of the simulation result. The MIP signal for the sensor with the lowest fluence is simulated for an electrical field of $E = 0.18\,\rm{V}/\mu\rm{m}$. The build-up of space charge leads to a two peak structure in the MIP pulse shape, which results in an overall reduced charge collection efficiency. The MIP particle for the highly irradiated diamond sample is simulated at an increased electrical field of $E = 0.72\,\rm{V}/\mu\rm{m}$. The build-up of space charge is as well influencing the signal shape of the MIP particle resulting in a reduced charge collection efficiency. Based on the simulated MIP signals, the calculated charge collection efficiency is shown in Figs.~\ref{Fig_EDM_CCE_1},d as function of exposure time to the $^{90}Sr$ source. The simulated charge collection efficiency is in agreement with the experimental measured CCE values.

\begin{figure}
\includegraphics*[width=\linewidth]{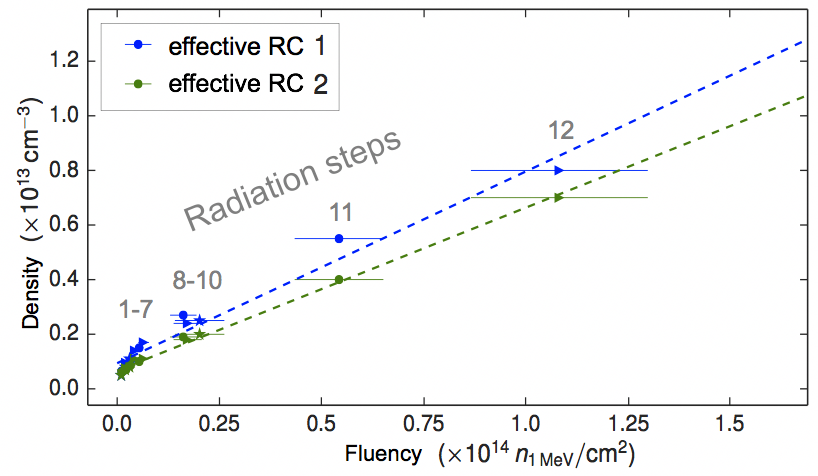}
\caption{%
  The optimized trap density of $eRC1$ (blue) and $eRC2$ (green) for the different irradiation damages of the proton and neutron irradiated diamond sensors. As expected, the trap densities of both effective recombination centers increase linearly with the irradiation damage.}
\label{SimulationvsMeasurement_1}
\end{figure}
\subsection{Effective defect model as function of radiation damage}
\begin{figure}
\includegraphics*[width=\linewidth]{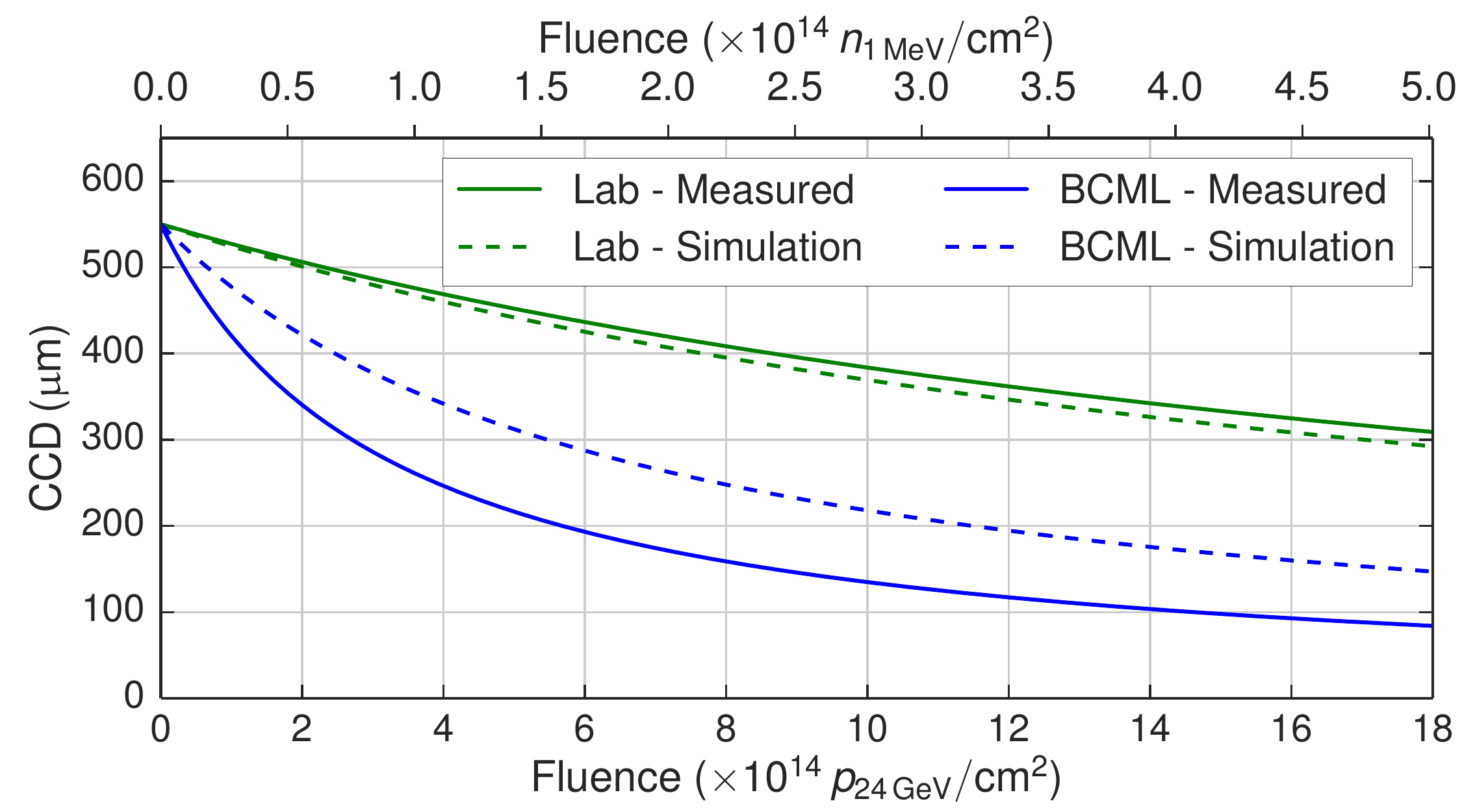}
\caption{%
Fits of the measured (solid) and simulated (dashed) CCD as function of fluence for an electrical field of ${E = 1.0\,\rm{V}/\mu\rm{m}}$ for the laboratory (green) and the BCML (blue) particle rate environment. The data points on which the fits are based rely on the measured and simulated radiation constants $k$ listed in Table \ref{Table_DynamicTrapModel_Sim_DiffEfields} and extracted from Figs.~\ref{BCML_signalloss},\ref{CCE_200V_2} and \ref{Fig_Chapter6_DynamicTrapModel_Sim_DiffEfields}.}
\label{CCDvsFluency}
\end{figure}

The optimization of the effective defect model to the experimental measurement results shown in Fig.~\ref{Fig_Chapter6_TCTPol_200V_Hole_SIM} and \ref{Fig_EDM_MIP_CCE} are representative for the fitting of the defect model for each of the 12 different irradiation steps. The optimized trap densities for both effective recombination centers $eRC1$ and $eRC2$ for each irradiation step are plotted as function of the radiation damage caused by the particle fluence of $\Phi\,(p_{\rm{24\,GeV\,eq.}}/\rm{cm}^{2})$ in Fig.~\ref{SimulationvsMeasurement_1}. The error to the absolute radiation damage in the x-dimension is due to the limited accuracy of the irradiation facilities. The accuracy of the proton irradiation is $\pm 20\%$ and the accuracy of the neutron irradiation is $\pm 30\%$. The linear regression of the trap densities with respect to the total radiation damage gives:
\begin{align}
\label{CCE_calc_eq}
\rho_{\rm{eRC1}} &= \Phi\cdot 0.0252\,(\rm{cm}^{-1})+ 9.40\times 10^{11}\,(\rm{cm}^{-3}),\\
\rho_{\rm{eRC2}} &= \Phi\cdot 0.0215\,(\rm{cm}^{-1}) + 6.67\times 10^{11}\,(\rm{cm}^{-3}),
\end{align}
with $\Phi$ as $(p_{\rm{24\,GeV\,eq.}}/\rm{cm}^{2})$ particle fluence. A possibility to verify this effective trap model is the simulation of the radiation induced signal degradation in terms of charge collection efficiency as function of radiation damage, see Fig.~\ref{CCDvsFluency}. Based on this degradation the radiation constant can be calculated using Eq.~\ref{Eq_RD42_fin} and be compared to the radiation constant found by the experimental measurements. The CCE of the diamond sensor is simulated for a typical particle rate environment created by a $^{90}Sr$ and at an electrical field of $E = 1\,\rm{V}/\mu\rm{m}$.

The simulated charge collection distances matches the experimental measurement results for the laboratory rate environment. Based on the simulation a radiation constant of $k_{\rm{sim.}} = (8.9 \pm 1.1) \times 10^{-19}\,\rm{cm}^2\mu\rm{m}^{-1}$ is calculated. The simulated radiation constant is in agreement with the experimental measurement result of $k_{\rm{meas.}} = 8.2\times 10^{-19}\,\rm{cm}^2\mu\rm{m}^{-1}$. This is however not surprising since the effective defect model was optimized to these particular measurements.

\begin{figure*}
\captionsetup[subfigure]{justification=centering}
\subfloat[Electric field]{%
\includegraphics*[width=0.3\textwidth]{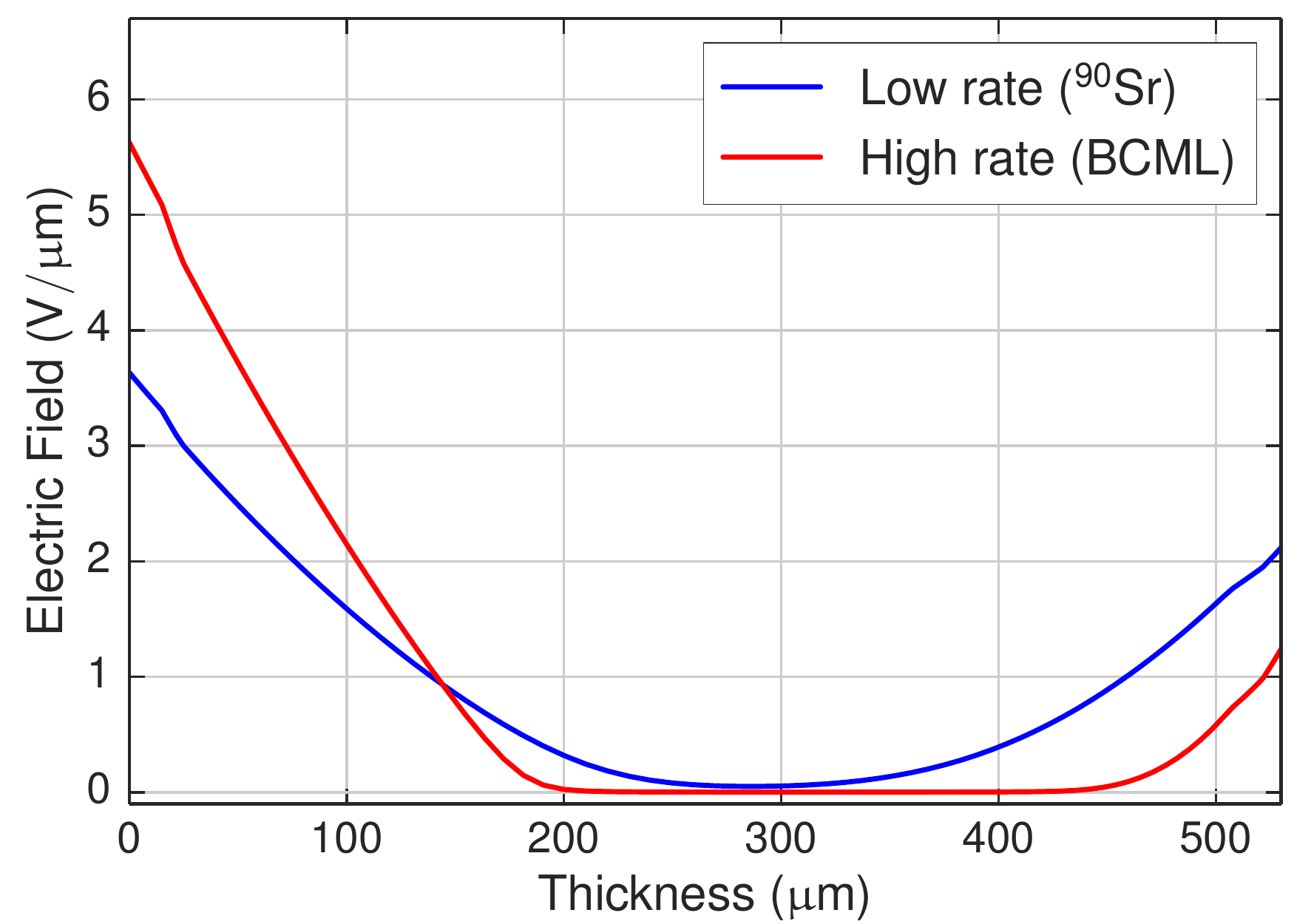}\label{Fig_EDM_ElectricProperties_Rate_variation_0}}\hfill
\subfloat[Space charge]{%
\includegraphics*[width=.3\textwidth]{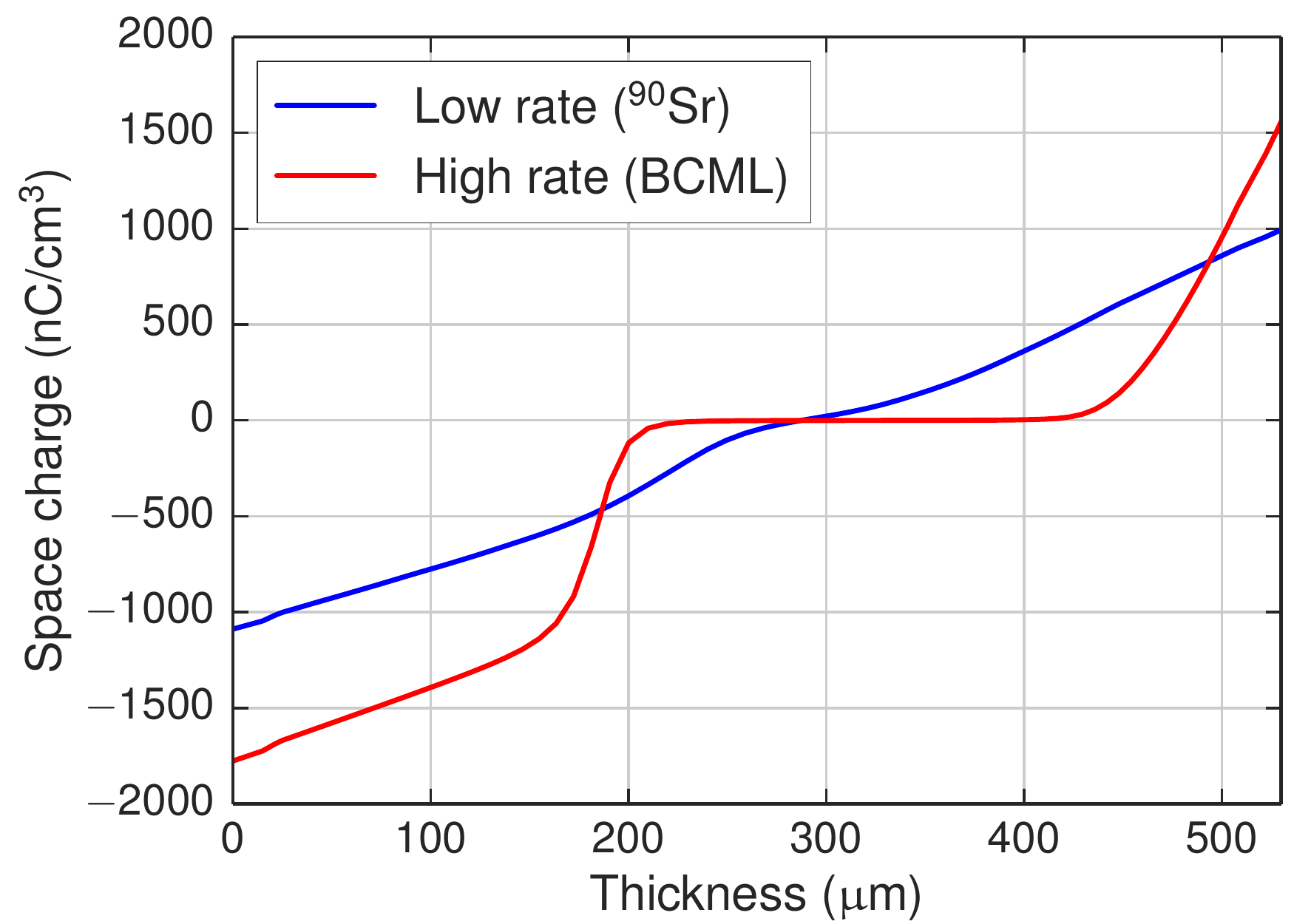}\label{Fig_EDM_ElectricProperties_Rate_variation_1}}\hfill
\subfloat[Normalized recombination]{%
\includegraphics*[width=.3\textwidth]{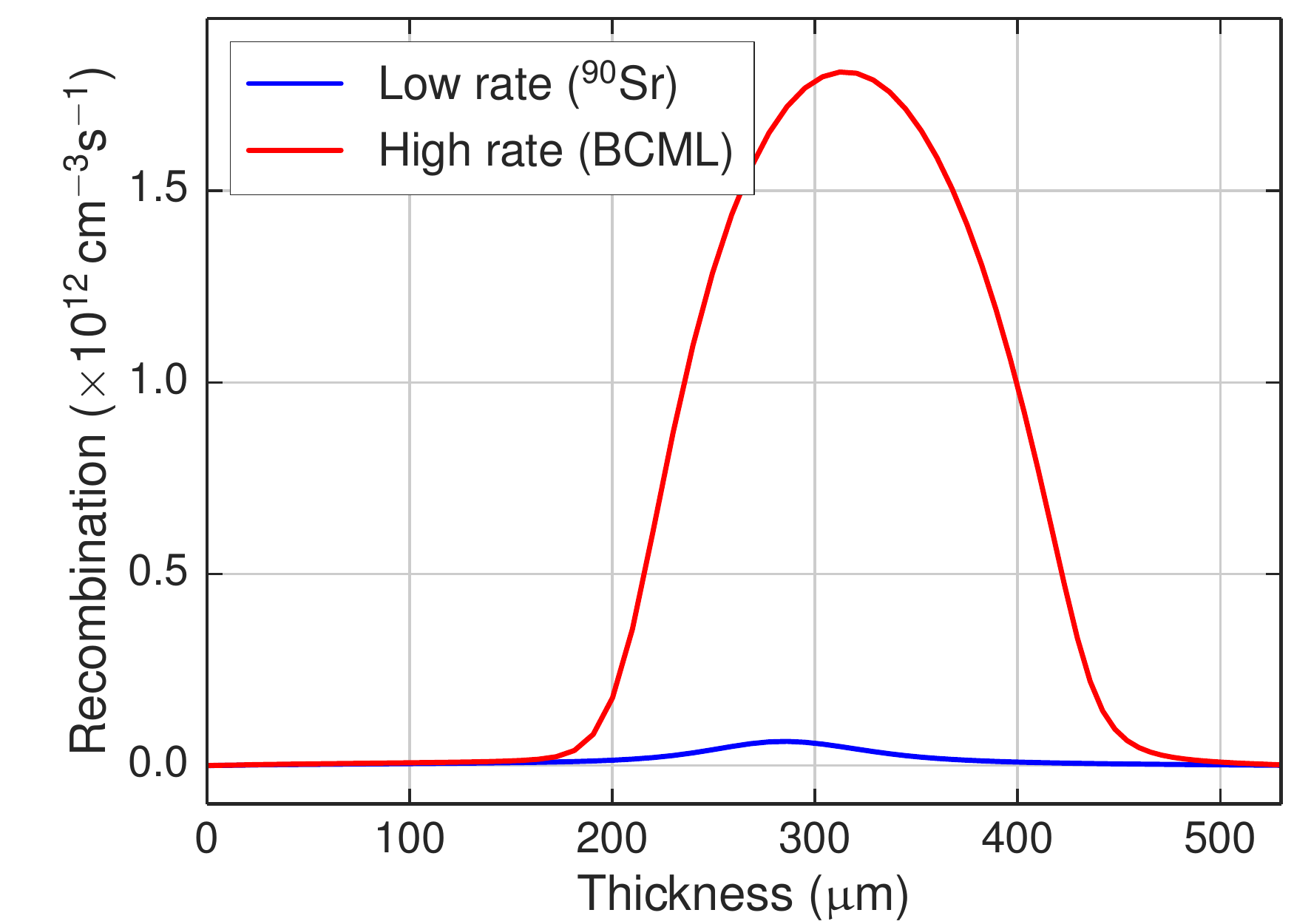}\label{Fig_EDM_ElectricProperties_Rate_variation_2}}\hfill
\caption{Simulation of diamond sensors operated at an external electrical field of $E = 1\,\rm{V}/\mu\rm{m}$ for a $^{90}Sr$ source (blue) and the BCML (red) particle rate environment.}
\label{Fig_EDM_ElectricProperties_Rate_variation}
\end{figure*}

\begin{figure}%
\centering
\includegraphics[width=\linewidth]{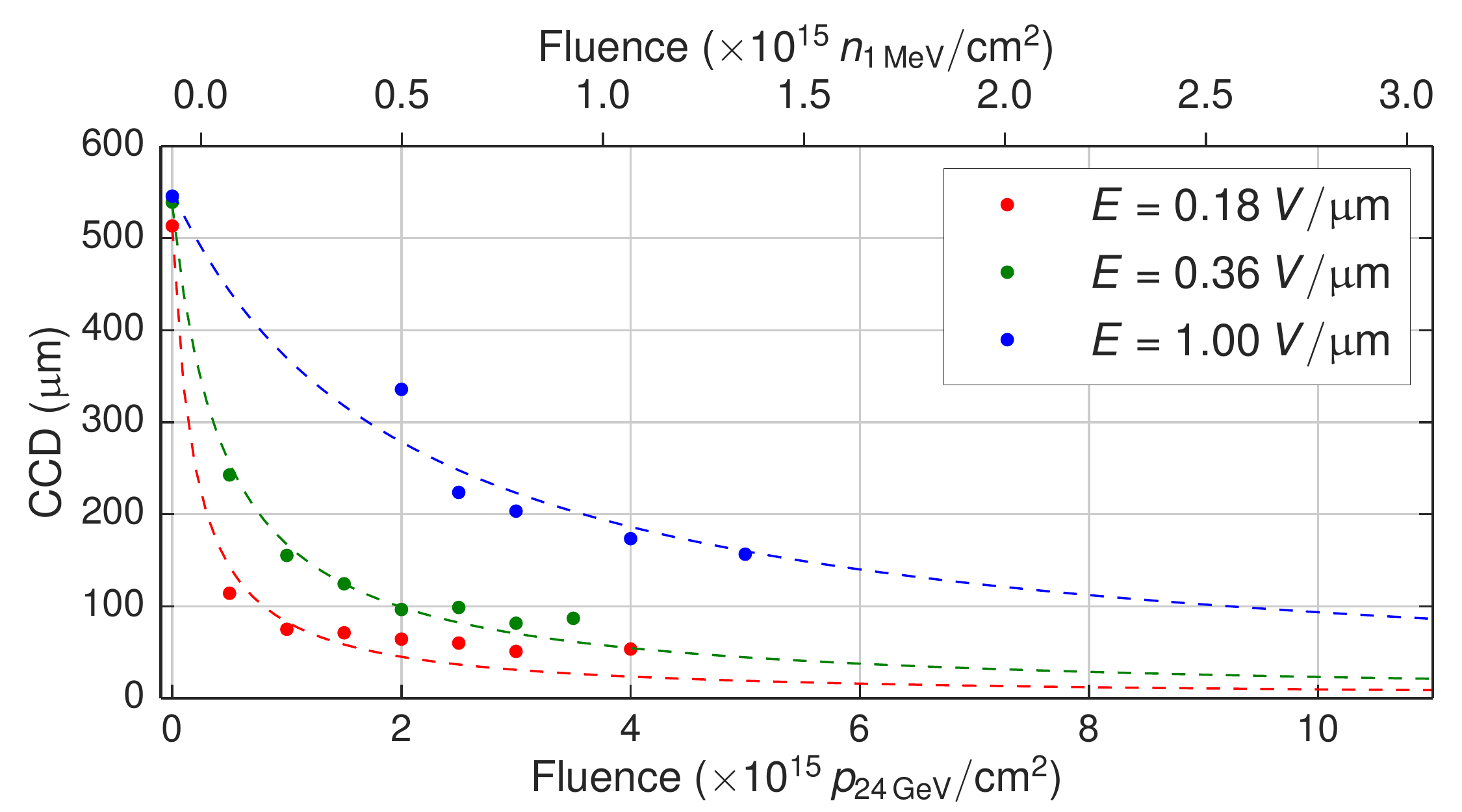}
\caption{Simulation of the CCD as function of fluence for a diamond sensor exposed to a $^{90}Sr$ source. The radiation induced signal degradation was simulated for three different electrical fields at which the diamond detectors are operated: $E = 0.18\,\rm{V}/\mu\rm{m}$ in red, $E = 0.36\,\rm{V}/\mu\rm{m}$ in green and $E = 1.00\,\rm{V}/\mu\rm{m}$ in blue. Based on the simulated CCD as function of radiation damage, the standard radiation model was fitted to the data, indicated in dashed lines.}
\label{Fig_Chapter6_DynamicTrapModel_Sim_DiffEfields}
\end{figure}

\begin{figure*}
\captionsetup[subfigure]{justification=centering}
\subfloat[Electric field]{%
\includegraphics*[width=0.3\textwidth]{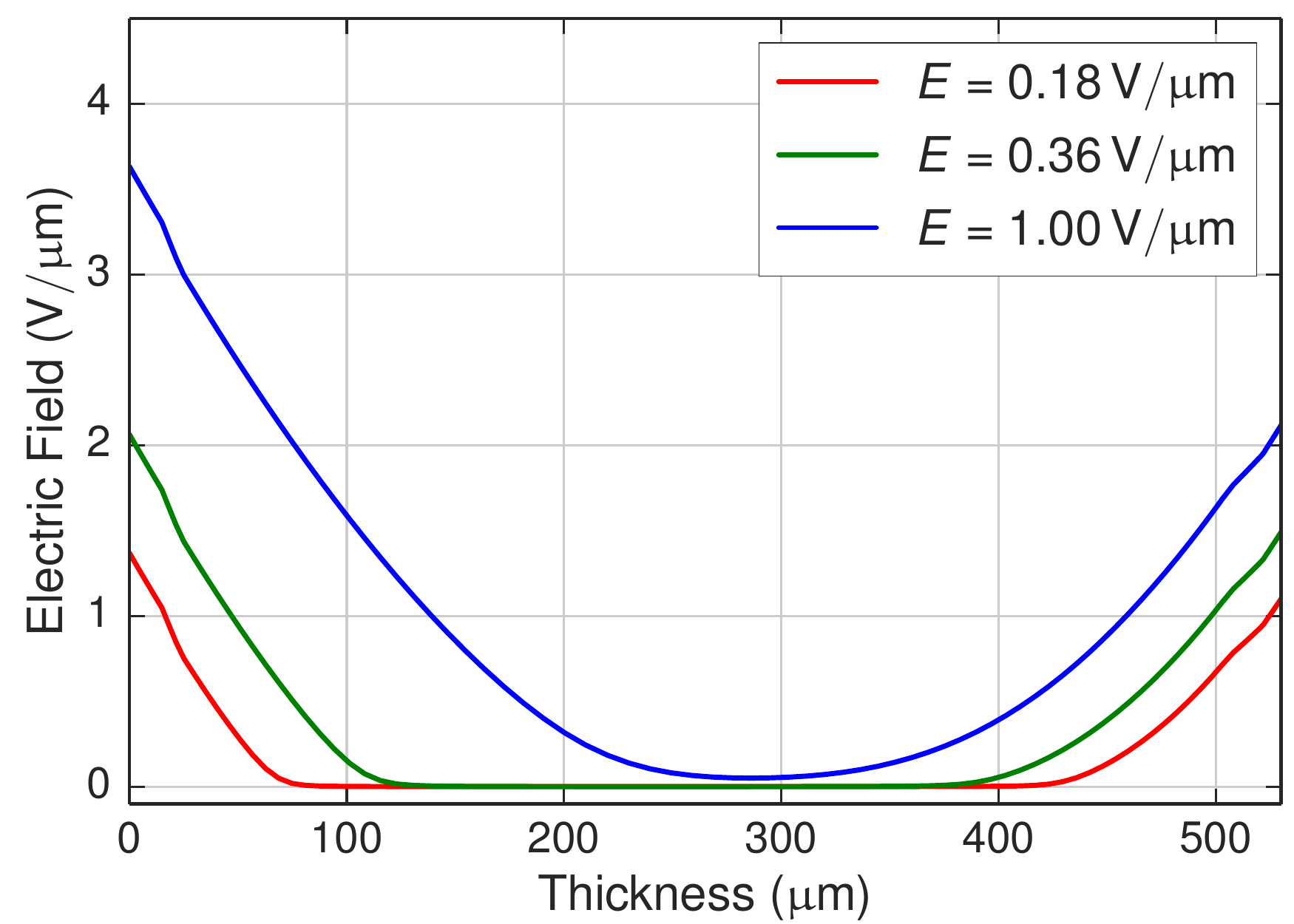}\label{Fig_Chapter6_DynamicTrapModel_Sim_DiffEfields_2_10_15_EFIELD}}\hfill
\subfloat[Space charge]{%
\includegraphics*[width=.3\textwidth]{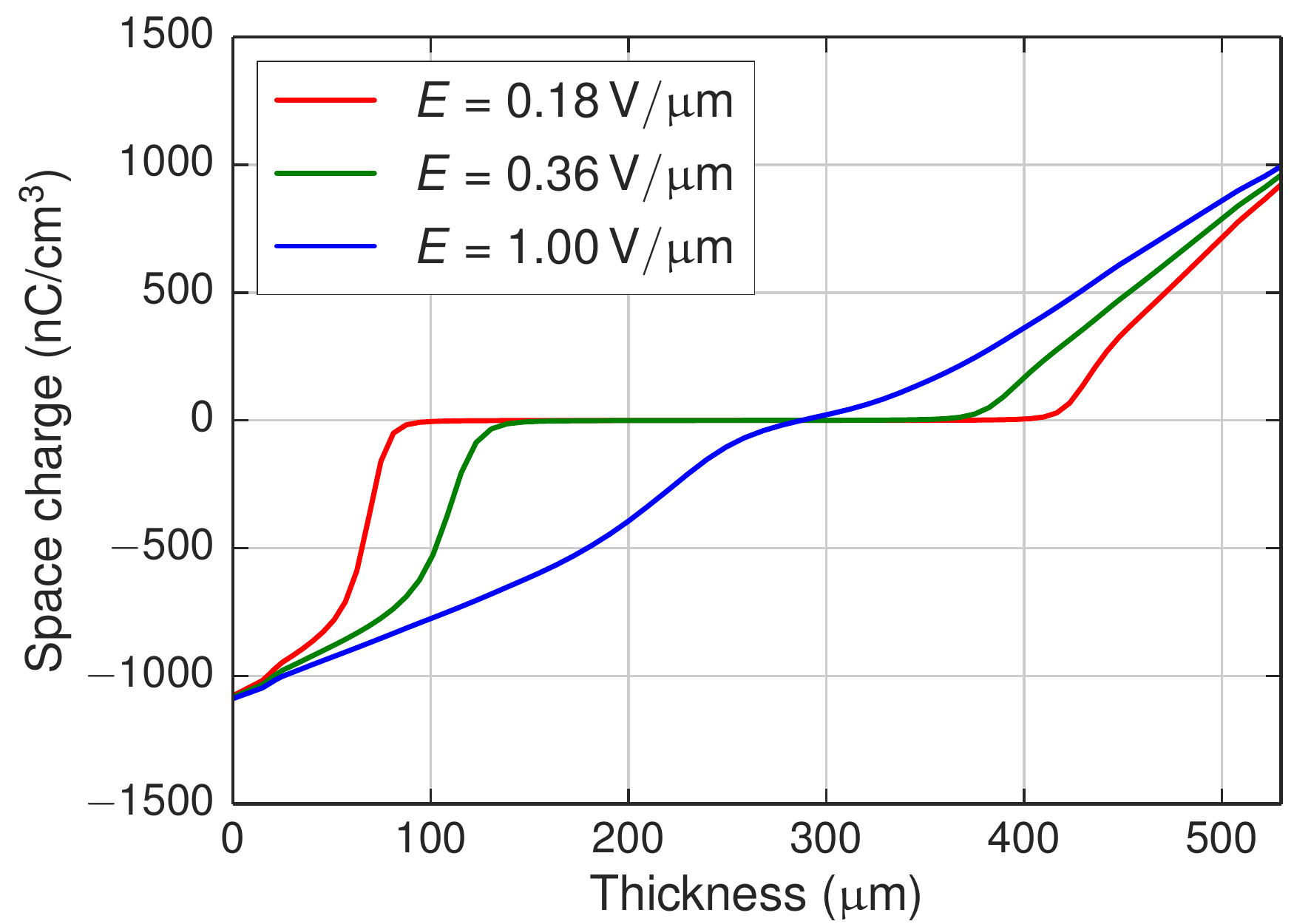}\label{Fig_Chapter6_DynamicTrapModel_Sim_DiffEfields_2_10_15_SpaceC}}\hfill
\subfloat[Normalized recombination]{%
\includegraphics*[width=.3\textwidth]{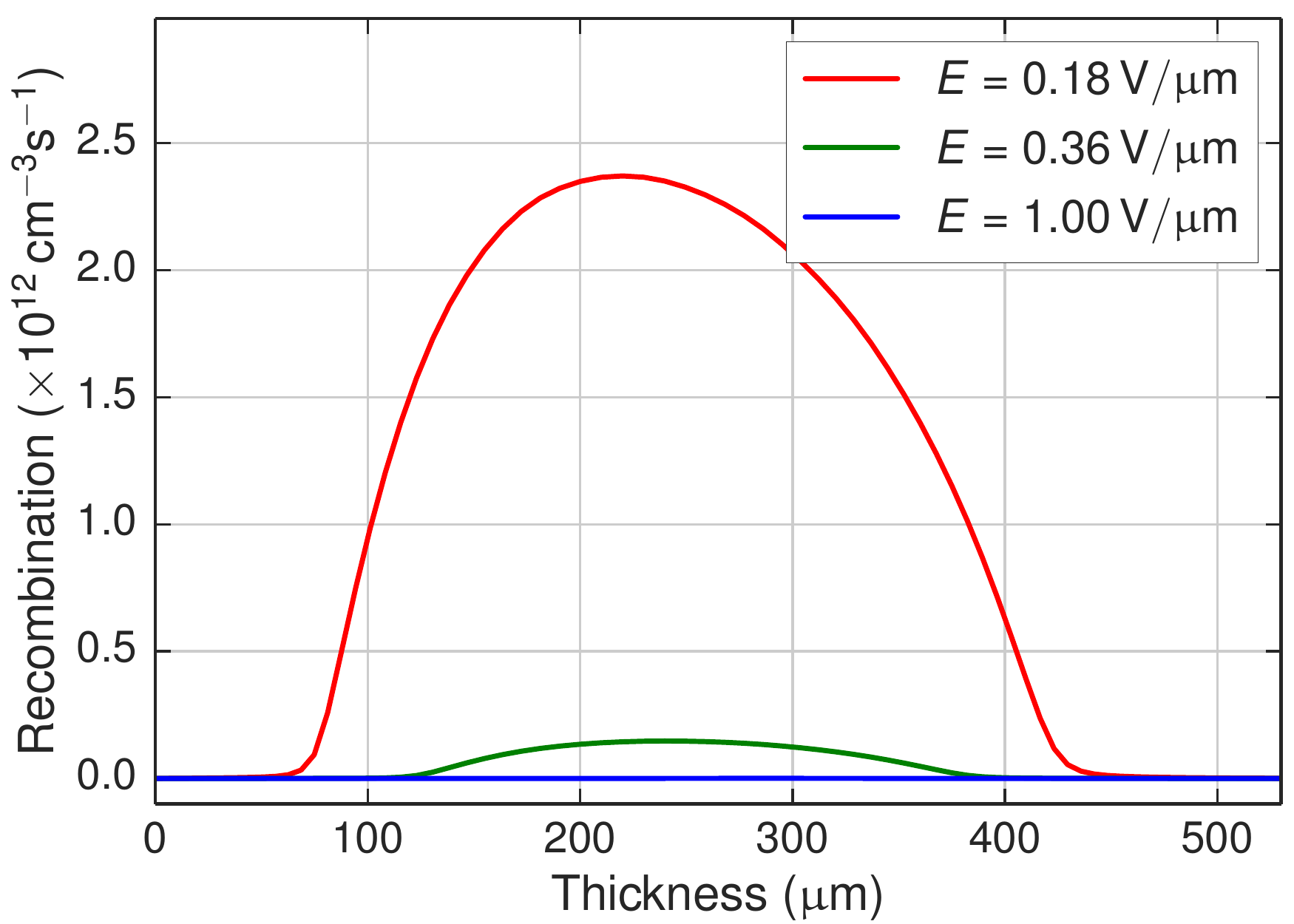}\label{Fig_Chapter6_DynamicTrapModel_Sim_DiffEfields_2_10_15_Recom}}\hfill
\caption{Simulation of damaged diamond sensors with a fluence of ${\Phi = 2 \times 10^{15}\,p_{\rm{24GeV}}/\rm{cm}^{2}}$ in a $^{90}Sr$ particle rate environment for three different electrical fields: $E = 1.00\,\rm{V}/\mu\rm{m}$ (blue), $E = 0.36\,\rm{V}/\mu\rm{m}$ (green) and $E = 0.18\,\rm{V}/\mu\rm{m}$ (red).}
\label{Fig_EDM_ElectricProperties_HV_variation}
\end{figure*}

The effective defect model is now used to simulate increased particle rate environments. In laboratory measurements the particle rate created by a $^{90}Sr$ source typically creates a particle rate of ${f_{^{90}Sr} = 0.15\,\rm{GHz}/\rm{cm}^3}$ \cite{Kassel2016}. The particle rates at the CMS detector are significantly higher and depend on the exact detector location. For the BCML detector location a MIP particle rate of $f_{\rm{BCML1}} \approx 10\,\rm{GHz}/\rm{cm}^3$ is estimated based on FLUKA \cite{FLUKA1,FLUKA2} simulations. The simulated charge collection distance for the increased particle rate is indicated in Fig.~\ref{CCDvsFluency} in blue dashed lines. The increased particle rate environment leads to a strongly reduced charged collection distance. After a radiation damage, corresponding to a particle fluence of ${\Phi = 10 \times 10^{14}\,p_{\rm{24GeV}}/\rm{cm}^{2}}$, the charge collection distance simulated for the particle rate environment at the BCML location is reduced by $53\%$ compared to the $^{90}Sr$ particle rate environment. Based on these simulation results a three times increased radiation constant of $k_{\rm{sim.}} = 27.7\times 10^{-19}\,\rm{cm}^2\mu\rm{m}^{-1}$ is calculated. This radiation constant can be directly compared to the radiation constant measured with the BCML pCVD diamond sensors at the CMS detector (indicated in solid blue), since the radiation constant is independent of the diamond material (sCVD or pCVD) used to measure the radiation induced signal degradation \cite{RD42_2008}. The different diamond material is only reflected in a different initial charge collection distance value, compare sec.~\ref{CCE_chapter}.

The calculated radiation constant based on the BCML detector degradation of $k_{\rm{meas.}} = 56.0\times 10^{-19}\,\rm{cm}^2\mu\rm{m}^{-1}$ is still by a factor of two higher than the simulation result, see Table~\ref{Table_DynamicTrapModel_Sim_DiffEfields}. This discrepancy can be caused e.g. by an underestimation of the radiation damage, which is based on a FLUKA simulation. Furthermore, the BCML sensors are exposed to a mixed particle field that could contribute to non linear degradation effects of the sensors. 

\begin{table}[t]%
\centering
  \caption{Overview of the measured and simulated radiation constants of diamond sensors for different particle rate environments. The diamond sensors were operated at an electrical field of ${E = 1.0\,\rm{V}/\mu\rm{m}}$.}
  \begin{tabular}[htbp]{@{}ccc@{}}
    \hline
    Particle rate & $k_{\rm{meas.}}$ & $k_{\rm{sim.}}$\\
     environment & ($\times 10^{-19}\,\rm{cm}^2/\mu\rm{m}$) & ($\times 10^{-19}\,\rm{cm}^2/\mu\rm{m}$)\\
    \hline
	$^{90}Sr$ & $8.2\pm 0.5$ & $8.9 \pm 1.1$\\
	CMS   & $56.0 \pm 4.1$& $27.7 \pm 0.2$\\
    \hline
  \end{tabular}
  \label{Table_DynamicTrapModel_Sim_DiffEfields}
\end{table}

A radiation damage of $\Phi = 1.5 \times 10^{15}\,p_{\rm{24GeV}}/\rm{cm}^{2}$ results in a signal degradation of 42\% and 70\% for the $^{90}Sr$ and BCML particle rate environment, respectively. The corresponding electrical field configurations of the diamond sensors operated at an electrical field of $E = 1\,\rm{V}/\mu\rm{m}$ are shown in Fig.~\ref{Fig_EDM_ElectricProperties_Rate_variation_0}. The expected electric field modification due to an almost linear build-up of space charge (Fig.~\ref{Fig_EDM_ElectricProperties_Rate_variation_1}) in the $^{90}Sr$ particle rate environment leads to a local minimum in the electric field. This results in a slightly increased recombination rate (Fig.~\ref{Fig_EDM_ElectricProperties_Rate_variation_2}) leading to the reduced CCE. 
The electrical field is significantly modified at the increased particle rate environment of the BCML detector location, see Fig.~\ref{Fig_EDM_ElectricProperties_Rate_variation_0} in red. The highly increased build-up of space charge leads to a suppression of the electrical field in about $\sim 230\,\mu\rm{m}$, that is about half of the sensor thickness. In this region the charge carrier recombination is strongly increased and explains the poor charge collection efficiency of $\sim 30\%$.

\section{Simulation of radiation damage for different electrical fields}
In this section the effective defect model is used to analyze the charge collection distance as function of the electric field at which the sensor is operated. In Fig.~\ref{Fig_Chapter6_DynamicTrapModel_Sim_DiffEfields} the radiation induced signal degradation is simulated for three different electric fields of $E = 1.00\,\rm{V}/\mu\rm{m}$, $E = 0.36\,\rm{V}/\mu\rm{m}$ and $E = 0.18\,\rm{V}/\mu\rm{m}$. Based on the simulation results the radiation constant is calculated and listed in Table~\ref{Table_Fig_Chapter6_DynamicTrapModel_Sim_DiffEfields}. Comparison the radiation constants demonstrates the importance of preferably high operational electric fields. A three times reduced electric field from 1 to 0.36\,$\rm{V}/\mu\rm{m}$ leads to an increased radiation constant by a factor of $\sim 4.7$. Furthermore, reducing the electric field by a factor of 5 to an electric field of $E = 0.18\,\rm{V}/\mu\rm{m}$ leads to a $\sim 11.3$ times increased radiation constant. Hence, the diamond polarization leads to a non-linear increase in the radiation induced signal degradation as function of the electric field at which the diamond is operated.

\begin{table}[t]%
\centering
  \caption{Overview of the simulated radiation constants of the diamond sensor operated at different electrical fields. The diamond was simulated for a particle rate environment created by a $^{90}Sr$ source.}
  \begin{tabular}[htbp]{@{}ccc@{}}
    \hline
    Electrical field $E$ & Radiation constant $k$\\
    ($\rm{V}/\mu\rm{m}$) & ($\times 10^{-19}\,\rm{cm}^2/\mu\rm{m}$)\\
    \hline
	1.00 & $8.9 \pm 1.1$\\
	0.36   & $41.5 \pm 2.6$\\
	0.18 & $101.0 \pm 15.7$\\
    \hline
  \end{tabular}
  \label{Table_Fig_Chapter6_DynamicTrapModel_Sim_DiffEfields}
\end{table}

A more detailed understanding of the diamond polarization leading to the severe signal degradation is obtained by analyzing the corresponding electric field and space charge inside the diamond sensors for a particular fluence, see Fig.~\ref{Fig_EDM_ElectricProperties_HV_variation}. Although the total amount of space charge is reduced for the lower electrical field configurations, the impact to the overall electric field inside the diamond sensor is strongly increased. This leads to the suppression of the electric field in more than half of the detector thickness leading to a strongly increased charge carrier recombination, see Fig.~\ref{Fig_Chapter6_DynamicTrapModel_Sim_DiffEfields_2_10_15_Recom}.

\section{Conclusion}
The radiation induced signal degradation of diamond detectors can be described using the effective defect model presented in this paper. This defect model was found by optimizing TCT and CCE simulations to the experimental measurement results of different irradiated diamond sensors. The simulation and measurement results underline the crucial role of the polarization effect to understand the radiation induced signal degradation of diamond detectors. The build-up of space charge leads to a locally reduced electric field at which the increased charge carrier recombination leads to a reduced CCE. Using the effective defect model to extrapolate the reduction of the electric field by the polarizing space charge inside the sensor to the high rate environment of the CMS detector explains the poor performance of the diamond sensors in this harsh environment of the LHC.

The effective defect model showed furthermore the importance of high bias voltages in order to minimize the radiation induced signal degradation. At reduced bias voltages the diamond polarization leads to a non-linear increase in the radiation induced signal degradation. Hence one should try to increase the high voltage breakthrough voltage, so one could operate with an electric field from the bias voltage well above the electric field from the space charge.
Alternatively, the switching of the bias voltage with a few Hz, so that space charge would switch direction as well and could not build-up inhomogeneously could avoid the diamond polarization.

\begin{acknowledgement}
This work has been sponsored by the Wolfgang Gentner Programme of the Federal Ministry of Education and Research and been supported by the H2020 project AIDA-2020, GA no. 654168 (http://aida2020.web.cern.ch/).
\end{acknowledgement}

\end{document}